\documentclass[prb,showpacs,amsmath,amssymb,amsfonts,superscriptaddress]{revtex4} 
\usepackage[T1]{fontenc}
\usepackage{bm}

\usepackage{times}
\usepackage{pslatex}
\usepackage{amsmath}
\usepackage{amsthm}
\usepackage{amssymb}
\usepackage{amsbsy}
\usepackage{graphicx}
\usepackage{pstricks}
\usepackage{wasysym}
\usepackage{mathrsfs}

\begin{document}

\newcommand{\hexaa}{\;\pspicture(0,0.1)(0.35,0.6)\psset{unit=0.75cm}
\pspolygon(0,0.15)(0,0.45)(0.2598,0.6)(0.5196,0.45)(0.5196,0.15)(0.2598,0)
\psset{linewidth=0.08,linestyle=solid}
\psline(0,0.15)(0,0.45)
\psline(0.2598,0.6)(0.5196,0.45)
\psline(0.5196,0.15)(0.2598,0)
\psdots[linecolor=gray,dotsize=.20](0,0.30)
\psdots[dotstyle=triangle*,linecolor=gray,dotsize=.20](0.3897,0.525)
\psdots[linecolor=gray,dotsize=.20,dotstyle=square*](0.3897,0.075)
\endpspicture\;}

\newcommand{\hexab}{\;
\pspicture(0,0.1)(0.35,0.6)
\psset{unit=0.75cm}
\pspolygon(0,0.15)(0,0.45)(0.2598,0.6)(0.5196,0.45)(0.5196,0.15)(0.2598,0)
\psset{linewidth=0.08,linestyle=solid}
\psline(0,0.15)(0,0.45)
\psline(0.2598,0.6)(0.5196,0.45)
\psline(0.5196,0.15)(0.2598,0)
\psdots[dotstyle=square*,linecolor=gray,dotsize=.20](0,0.30)
\psdots[linecolor=gray,dotsize=.20](0.3897,0.525)
\psdots[dotstyle=triangle*,linecolor=gray,dotsize=.20](0.3897,0.075)
\endpspicture\;}

\newcommand{\hexb}{\;
\pspicture(0,0.1)(0.35,0.6)
\psset{unit=0.75cm}
\psset {linewidth=0.03,linestyle=solid}
\pspolygon[](0,0.15)(0,0.45)(0.2598,0.6)(0.5196,0.45)(0.5196,0.15)(0.2598,0)
\psset{linewidth=0.08,linestyle=solid}
\psline(0.2598,0.6)(0,0.45)
\psline(0.5196,0.15)(0.5196,0.45)
\psline(0.2598,0)(0,0.15)
\psdots[linecolor=gray,dotsize=.20](0.1299,0.525)
\psdots[dotstyle=triangle*,linecolor=gray,dotsize=.20](0.525,0.3)
\psdots[linecolor=gray,dotsize=.20,dotstyle=square*](0.1299,0.075)
\endpspicture\;}

\newcommand{\hexa}{ \; \pspicture(0,0.1)(0.6,0.6)
  \psdots[linecolor=black,dotsize=.15](0,0.30)(0.3897,0.525) (0.3897,0.075)
  \psset{linewidth=0.03,linestyle=solid}
  \pspolygon[](0,0.15)(0,0.45)(0.2598,0.6)(0.5196,0.45)(0.5196,0.15)(0.2598,0)
  \psset{linewidth=0.03,linestyle=solid} \psset{linewidth=0.08,linestyle=solid}
  \psset{linewidth=0.08,linestyle=solid}
  \psline[linecolor=black](0,0.15)(0,0.45)
  \psline[linecolor=black](0.2598,0.6)(0.5196,0.45)
  \psline[linecolor=black](0.5196,0.15)(0.2598,0) \endpspicture\;}

\newcommand{\hexaf}{ \; \psset{unit=0.7cm} \pspicture(0,0.1)(1,0.5)
  \psset{linewidth=0.03,linestyle=solid}
  \pspolygon[](0,0.25)(0.25,0)(0.75,0)(1,0.25) (0.75, 0.5) (0.25, 0.5)
  \psset{linewidth=0.03,linestyle=solid} \pspolygon[fillcolor=lightgray,
  fillstyle=solid](0.25,0.0)(0.75,0)(0.5,0.25) \pspolygon[fillcolor=lightgray,
  fillstyle=solid](0.25,0.5)(0.75,0.5)(0.5,0.25) \psline[](0.5,0)(0.5,0.5)
  \psset{linewidth=0.08,linestyle=solid}
  \psdots[linecolor=black,dotsize=.15](0.25,0)(0.25,0.5) (1,0.25)(0.5,0.25)  
\endpspicture\;}

\newcommand{\hexbf}{ \; \psset{unit=0.7cm} \pspicture(0,0.1)(1,0.5)
  \psset{linewidth=0.03,linestyle=solid}
  \pspolygon[](0,0.25)(0.25,0)(0.75,0)(1,0.25) (0.75, 0.5) (0.25, 0.5)
  \psset{linewidth=0.03,linestyle=solid} \pspolygon[fillcolor=lightgray,
  fillstyle=solid](0.25,0.0)(0.75,0)(0.5,0.25) \pspolygon[fillcolor=lightgray,
  fillstyle=solid](0.25,0.5)(0.75,0.5)(0.5,0.25) \psline[](0.5,0)(0.5,0.5)
  \psset{linewidth=0.08,linestyle=solid}
  \psdots[linecolor=black,dotsize=.15](0,0.25)(0.75,0) (0.75,0.5)(0.5,0.25) 
\endpspicture\;}

\newcommand{\hexafg}{ \; \psset{unit=0.7cm} \pspicture(0,0.1)(1,0.5)
  \psset{linewidth=0.03,linestyle=solid}
  \pspolygon[](0,0.25)(0.25,0)(0.75,0)(1,0.25) (0.75, 0.5) (0.25, 0.5)
  \psset{linewidth=0.03,linestyle=solid} \pspolygon[fillcolor=lightgray,
  fillstyle=solid](0.25,0.0)(0.75,0)(0.5,0.25) \pspolygon[fillcolor=lightgray,
  fillstyle=solid](0.25,0.5)(0.75,0.5)(0.5,0.25) \psline[](0.5,0)(0.5,0.5)
  \psset{linewidth=0.08,linestyle=solid}
  \psdots[linecolor=black,dotsize=.15](0.25,0)(0.25,0.5)(1,0.25)
  \psdots[linecolor=gray,dotsize=.15](0.5,0.25)  
\endpspicture\;}

\newcommand{\hexafo}{ \; \psset{unit=0.7cm} \pspicture(0,0.1)(1,0.5)
  \psset{linewidth=0.03,linestyle=solid}
  \pspolygon[](0,0.25)(0.25,0)(0.75,0)(1,0.25) (0.75, 0.5) (0.25, 0.5)
  \psset{linewidth=0.03,linestyle=solid} \pspolygon[fillcolor=lightgray,
  fillstyle=solid](0.25,0.0)(0.75,0)(0.5,0.25) \pspolygon[fillcolor=lightgray,
  fillstyle=solid](0.25,0.5)(0.75,0.5)(0.5,0.25) \psline[](0.5,0)(0.5,0.5)
  \psset{linewidth=0.08,linestyle=solid}
  \psdots[linecolor=black,dotsize=.15](0.25,0)(0.25,0.5)(1,0.25)
\endpspicture\;} 

\newcommand{\hexbfg}{ \; \psset{unit=0.7cm} \pspicture(0,0.1)(1,0.5)
  \psset{linewidth=0.03,linestyle=solid}
  \pspolygon[](0,0.25)(0.25,0)(0.75,0)(1,0.25) (0.75, 0.5) (0.25, 0.5)
  \psset{linewidth=0.03,linestyle=solid} \pspolygon[fillcolor=lightgray,
  fillstyle=solid](0.25,0.0)(0.75,0)(0.5,0.25) \pspolygon[fillcolor=lightgray,
  fillstyle=solid](0.25,0.5)(0.75,0.5)(0.5,0.25) \psline[](0.5,0)(0.5,0.5)
  \psset{linewidth=0.08,linestyle=solid}
  \psdots[linecolor=black,dotsize=.15](0,0.25)(0.75,0) (0.75,0.5)
  \psdots[linecolor=gray,dotsize=.15](0.5,0.25)  
\endpspicture\;}

\newcommand{\hexbfo}{ \; \psset{unit=0.7cm} \pspicture(0,0.1)(1,0.5)
  \psset{linewidth=0.03,linestyle=solid}
  \pspolygon[](0,0.25)(0.25,0)(0.75,0)(1,0.25) (0.75, 0.5) (0.25, 0.5)
  \psset{linewidth=0.03,linestyle=solid} \pspolygon[fillcolor=lightgray,
  fillstyle=solid](0.25,0.0)(0.75,0)(0.5,0.25) \pspolygon[fillcolor=lightgray,
  fillstyle=solid](0.25,0.5)(0.75,0.5)(0.5,0.25) \psline[](0.5,0)(0.5,0.5)
  \psset{linewidth=0.08,linestyle=solid}
  \psdots[linecolor=black,dotsize=.15](0,0.25)(0.75,0) (0.75,0.5)
\endpspicture\;}

\newcommand{\smallhexv}{\; \psset{unit=0.7cm} \pspicture(0,0.1)(0.2,0.35)
  \psset{linewidth=0.03,linestyle=solid}
  \pspolygon[](0.1,-0.1)(0,0.0)(0,0.2)(0.1,0.3)(0.2,0.20)(0.2,-0.0) 
\endpspicture \;}

\newcommand{\smallhexh}{ \psset{unit=0.7cm} \pspicture(0,0.1)(0.4,0.2)
  \psset{linewidth=0.03,linestyle=solid}
  \pspolygon[](0,0.1)(0.1,0.0)(0.3,0)(0.4,0.1)(0.3,0.2)(0.1,0.2)  
\endpspicture\;}

\newcommand{\smallrech}{ \; \pspicture(0,0.1)(0.2,0.1)
\psset{linewidth=0.025,linestyle=solid}
\psline[](0.1,0)(0.1,0.1)
\pspolygon[](0,0)(0.2,0)(0.2,0.1)(0,0.1)
\endpspicture\;}

\newcommand{\smallrecv}{ \; \pspicture(0,0.1)(0.1,0.2)
\psset{linewidth=0.025,linestyle=solid}
\psline[](0,0.1)(0.1,0.1)
\pspolygon[](0,0)(0,0.2)(0.1,0.2)(0.1,0.0)
\endpspicture\;}


\title{Strongly correlated electrons on frustrated lattices}

\author{Peter Fulde}
\address{Max-Planck-Institut f{\"u}r Physik komplexer Systeme, 01187 Dresden,
  Germany} 
\address{Asia Pacific Center for Theoretical Physics, Pohang, Korea}
\author{Frank Pollmann}
\address{Max-Planck-Institut f{\"u}r Physik komplexer Systeme, 01187 Dresden,
  Germany} 
\author{Erich Runge}
\address{Technische Universit\"at Ilmenau, 98683 Ilmenau, Germany} 

\date{\today}

\begin{abstract}
We give an overview of recent work on charge degrees of freedom of strongly
correlated electrons on geometrically frustrated lattices. Special attention is
paid to the checkerboard lattice, i.e., the two-dimensional version of a
pyrochlore lattice and to the kagome lattice. For the checkerboard lattice it
is shown that at half filling when spin degrees of freedom are neglected and
quarter filling when they are included excitations with fractional charges
$\pm$e/2 may exist. The same holds true for the three-dimensional pyrochlore
lattice. In the former case the fractional charges are confined. The origin of
the weak constant confining force is discussed and some similarities to quarks
and to string theory are pointed out. For the checkerboard lattice a
formulation in terms of a compact U(1) gauge theory is described. Furthermore a
new kinetic mechanism for ferromagnetism at special fillings of a kagome
lattice is discussed.\\ 
\end{abstract}

\maketitle



\section{Introduction}

\label{Sect:Introduction}

Charge is quantized in nature. Even subparticles like quarks which can't exist
separately (quark confinement) carry a quantized charge. It
came as a surprise when Su, Schrieffer and Heeger [\onlinecite{Su79,Su81}]
pointed out that in trans-polyacetylene (CH)$_n$ excitations may exist that
carry only a rational fraction of the electronic charge $e$, provided those
polymer chains are properly doped. Since then we have got used to the fact that
in solids fractionally charged excitations may exist which are either
deconfined or confined  [\onlinecite{Laughlin83,hou07,FuldeP02}]. It has also
become clear that two different types of systems with fractionally charged
excitations may occur. One type to which trans-polyacetylene or graphene
[\onlinecite{hou07}] belong does not require electron interactions for the
existence of fractional charges. However lattice degrees of freedom must be
included in order that they may form. For the second type of systems electron
interactions are crucial for the occurrence of fractional charges. The 
Fractional Quantum Hall Effect (FQHE) [\onlinecite{Laughlin83}] and strongly
correlated electrons on geometrically frustrated lattices
[\onlinecite{FuldeP02}] are examples of that category. 

Trans-polyacetylene, graphene and the FQHE are low dimensional
systems. Therefore the question arises whether fractionally charged
excitations may also appear in three dimensions. Indeed, in 2002 it was
suggested by one of us [\onlinecite{FuldeP02}] that excitations with charge
$\pm$e/2 may exist in pyrochlore and other geometrically frustrated lattices
and therefore also in three dimensions. This is of interest since in two
dimensional systems like the FQH liquid or graphene excitations with fractional
charges obey fractional statistics [\onlinecite{wilc90}]. They are neither
fermions nor bosons but instead anyons. If fractionally charged excitations
would always imply fractional or anyonic statistics then one would exclude such
excitations in three dimensional systems the reason being that in 3D only
fermions or bosons can exist as free particles. The possible existence of fractionally charged
excitations in a pyrochlore lattice excludes a simple one to one correspondence
between fractional charges and fractional statistics.

\noindent We argue that a general prerequisite for fractional charges based on
electron interactions are strong short-range correlations and special lattice
fillings. Following the general usage, we will simply call a lattice a
''frustrated'' one when the short-range electronic correlations are
incompatible with the lattice structure. It would be more appropriate to speak
of ''interactions, frustrated by the lattice'' but that is not customary.

\noindent Here we want to give an overview of recent studies which have been
made on charge degrees of freedom of strongly correlated spinless (or fully
spin polarized) fermions on frustrated lattices
[\onlinecite{FuldeP02,Runge04,Pollmann06,Pollmann06a,Pollmann06b,Pollmann06c,pollmann07b,pollmann07d}]. 
They refer mainly to the pyrochlore lattice and its two dimensional projection,
i.e., the checkerboard lattice. But also the kagome lattice will be
considered in special cases [\onlinecite{pollmann07e}]. The work on the
pyrochlore lattice has been stimulated by experimental findings on LiV$_2$O$_4$
a transition metal spinel [\onlinecite{Kondo97}]. It was found that it exhibits
heavy quasiparticle behavior, a hallmark of strongly correlated electron
systems. Note that spinels are of the composition AB$_2$O$_4$ with the B sites
forming a pyrochlore lattice.


\section{Charges on frustrated lattices}

\label{Sect:ChargFrustLatt}

\begin{figure}
\begin{center}
\begin{tabular}{ccc}
(a)~\includegraphics[height=25mm,keepaspectratio]{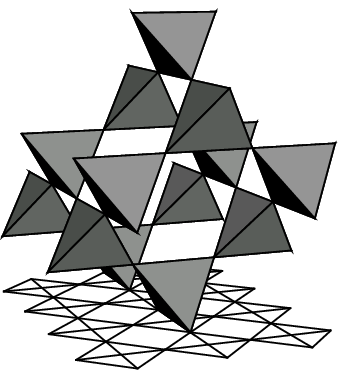}&
(b)~\includegraphics[height=25mm,keepaspectratio]{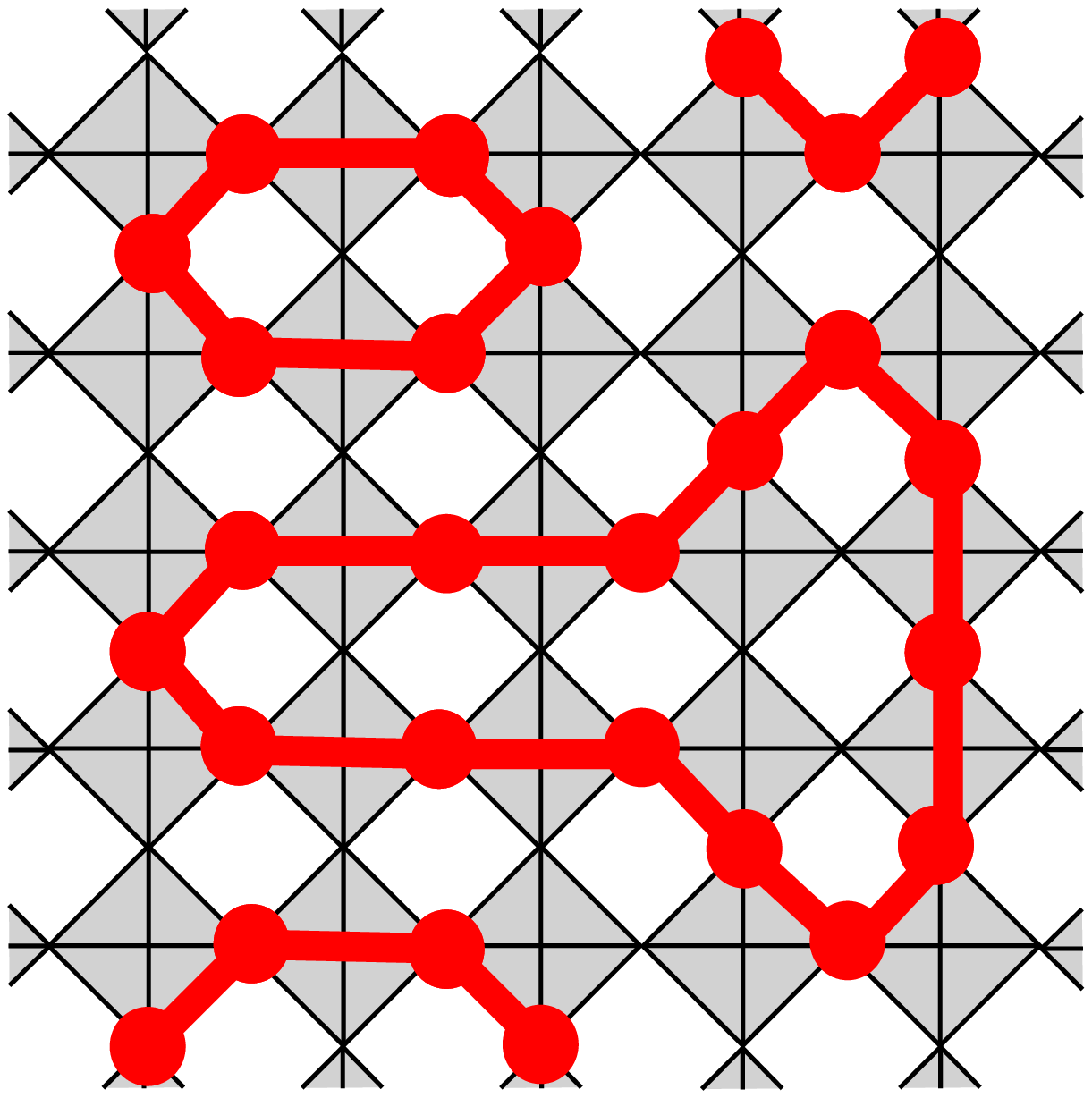}
&
(c)~\includegraphics[clip,height=25mm,keepaspectratio]{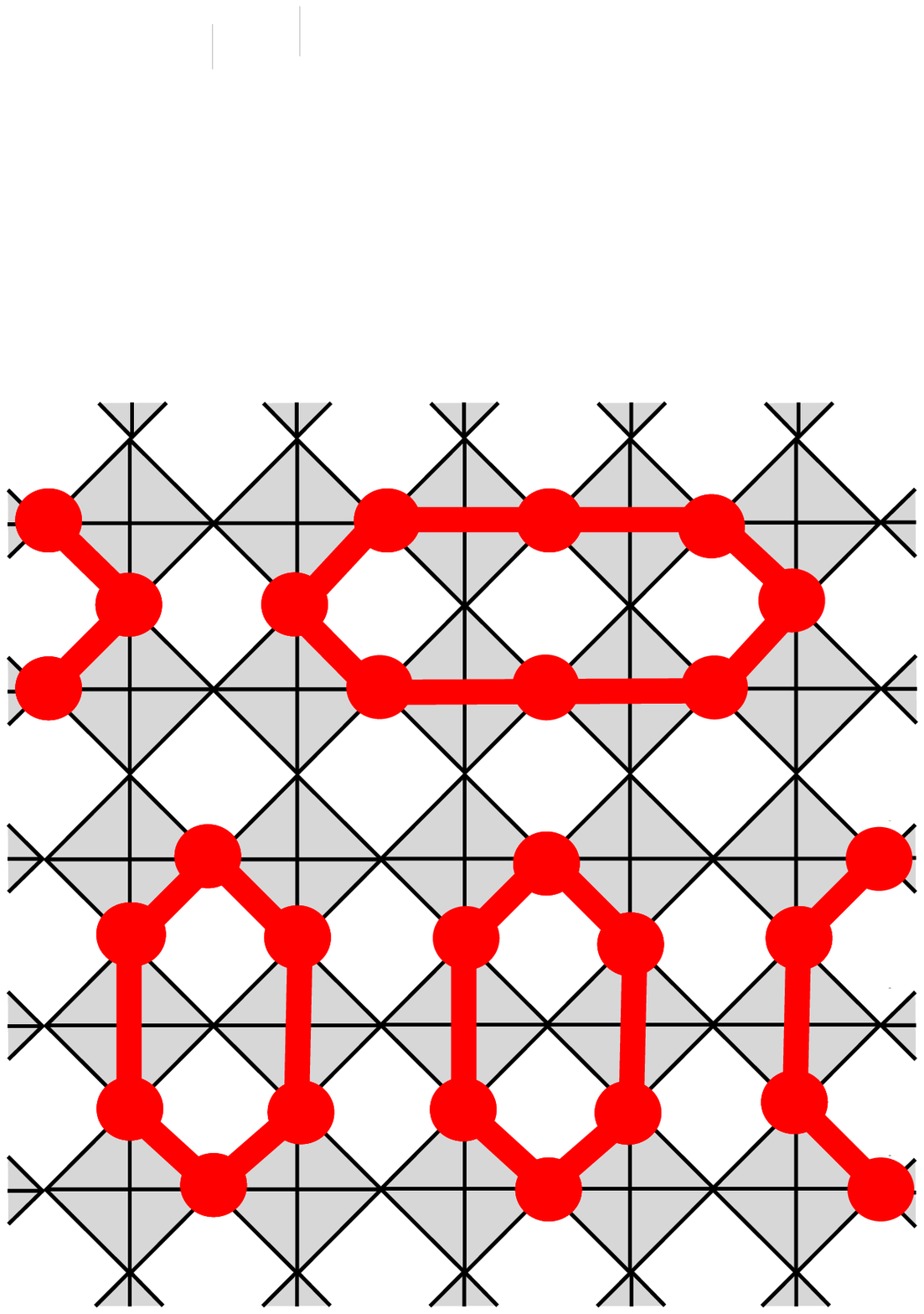}
\end{tabular}
\end{center}
\caption{(a) Checkerboard lattice as 2D projection of the 3D pyrochlore
  lattice [\onlinecite{Moessner04a}].~~~\label{cap:C2_pyro_checker}(b), (c) Two
  examples of allowed configurations on a checkerboard lattice at half filling.
Occupied sites are connected by thick solid lines as guides to the eye.
\label{cap:C2_allowed}}
\end{figure}

In the following we want to concentrate on the crisscrossed checkerboard
lattice because a number of features we want to point out are more easily
visualized on that lattice than on the pyrochlore lattice. The
checkerboard lattice can be considered as a 2D projection of the 3D pyrochlore
lattice (see FIG. \ref{cap:C2_pyro_checker}(a), and we treat lattice points
connected by a line as nearest neighbors. Since we are interested in charge
degrees of freedom, we disregard the spin. The following Hamiltonian for fully
spin polarized electrons or alternatively spinless fermions is assumed to hold

\begin{equation}
H = -t \sum_{\langle ij \rangle} \left( c^{\dag}_i c^{\vphantom{\dag}}_j + H.c. \right) + V
\sum_{\langle ij \rangle} n_i n_j~~~.
\label{eq1}
\end{equation} 

\noindent  The operators $c^{\dag}_i$ create fermions on sites $i$. The density operators are
$n_i = c^{\dag}_i c_i^{\vphantom{\dag}}$. We assume a system of $N$ sites filled with $N/2 =
\sum\limits_i n_i$ fermions (half filling) and focus on the strong correlation regime, i.e, $|t| \ll V$. 
Without loss of generality we assume $t>0$.

\subsection{Ground-state degeneracy:}

Consider first the case of $t = 0$. The nearest-neighbor repulsions $V$
are minimized if on each crisscrossed square two of the sites are occupied
while two sites remain empty. This we shall call the tetrahedron rule since it
applies equally well to the tetrahedra of the pyrochlore structure
[\onlinecite{Anderson56}]. All other configurations have a larger potential
energy. This implies immediately that the ground state is macroscopically
degenerate. More precisely the degeneracy is $N_{\rm deg} = \left( 4/3
\right)^{\frac{3}{4}N}$ and therefore the same as in the two-dimensional ice
model [\onlinecite{Lieb67}]. Two of those configurations are shown 
in FIGs. \ref{cap:C2_allowed}(b,c) where neighboring occupied sites have been
connected by a solid line. Note that a related Klein type spin model on the checkerboard lattice has been studied in Ref. [\onlinecite{Nussinov2007}] and shown to have a similar ground-state degeneracy. It is noticed that each configuration obeying the
tetrahedron rule, which we call {\it allowed} configurations in the following
causes a complete loop covering of the plane. When dynamics is added to the
system the time evolution of the loops will give raise to world sheets in a
space-time continuum instead of world lines as single particle propagation
does. One may speak therefore of a simple form of string theory which is
realized here. For an extended discussion of that issue see Section
\ref{Sect:StringsTimeEv}.

\begin{figure}
\begin{center}
\begin{tabular}{ccc}
(a)~\includegraphics[height=22mm,keepaspectratio]
{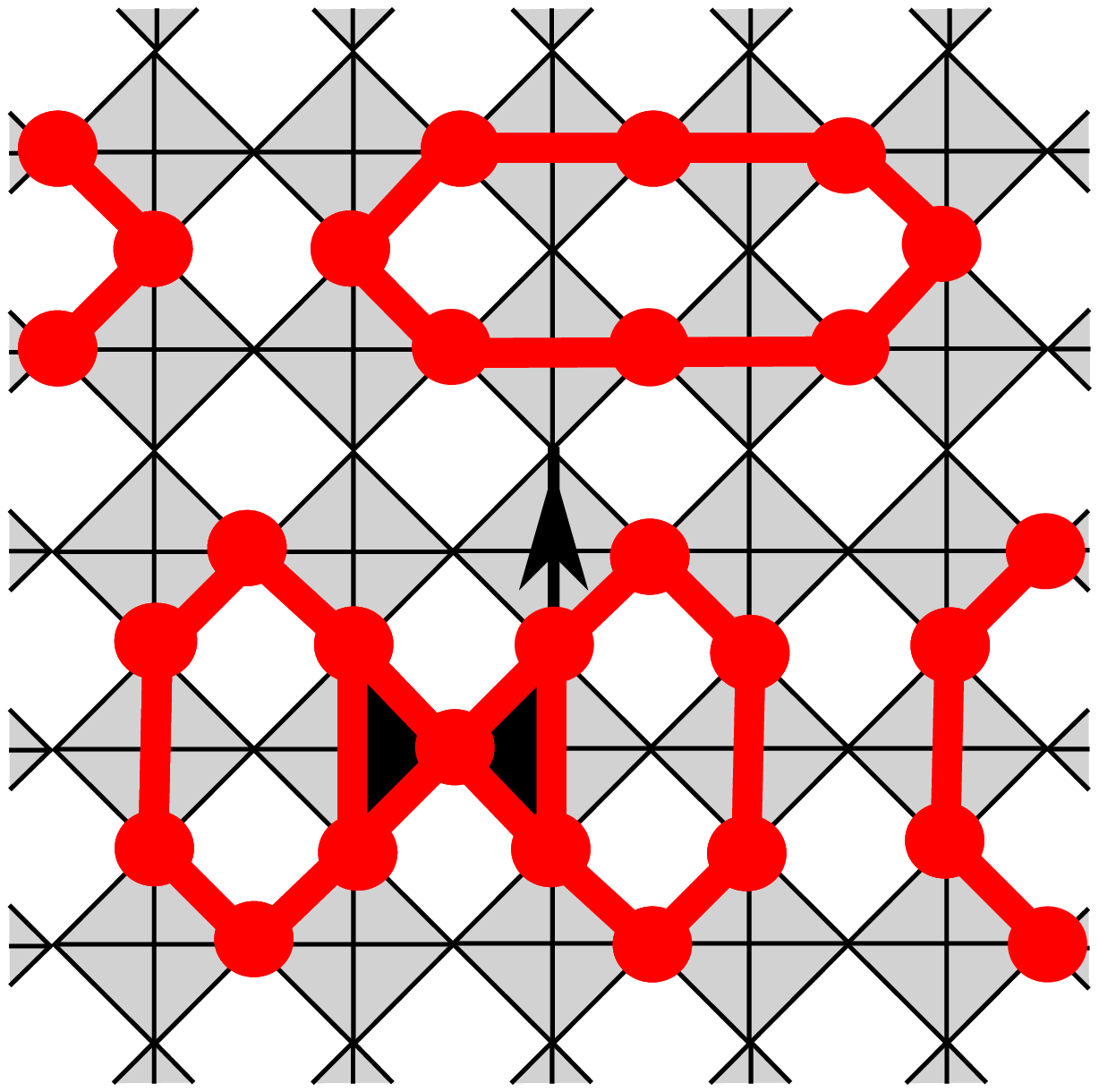}
&
(b)~\includegraphics[clip,height=22mm,keepaspectratio]
{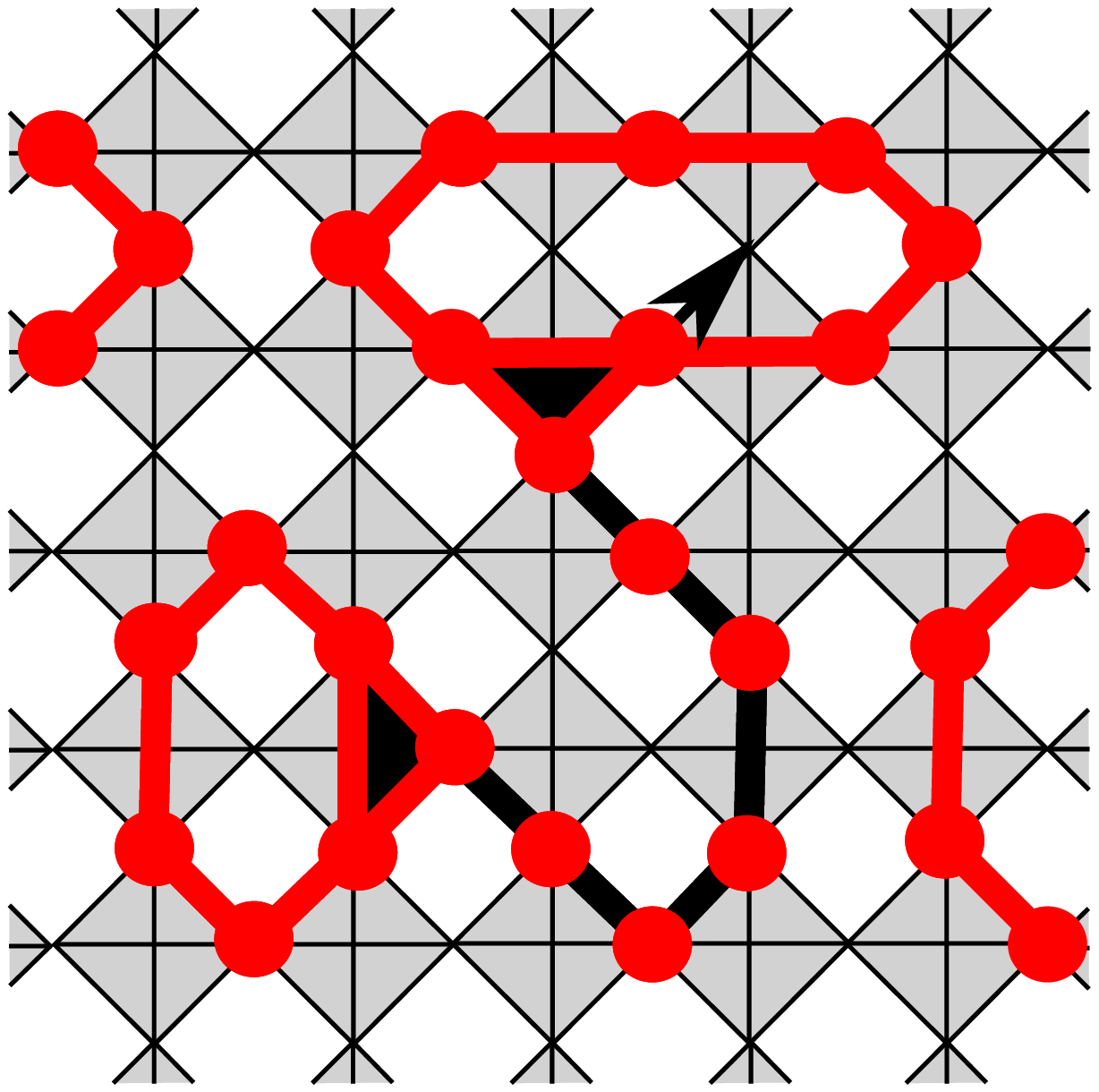}
&
(c)~\includegraphics[clip,height=22mm,keepaspectratio]
{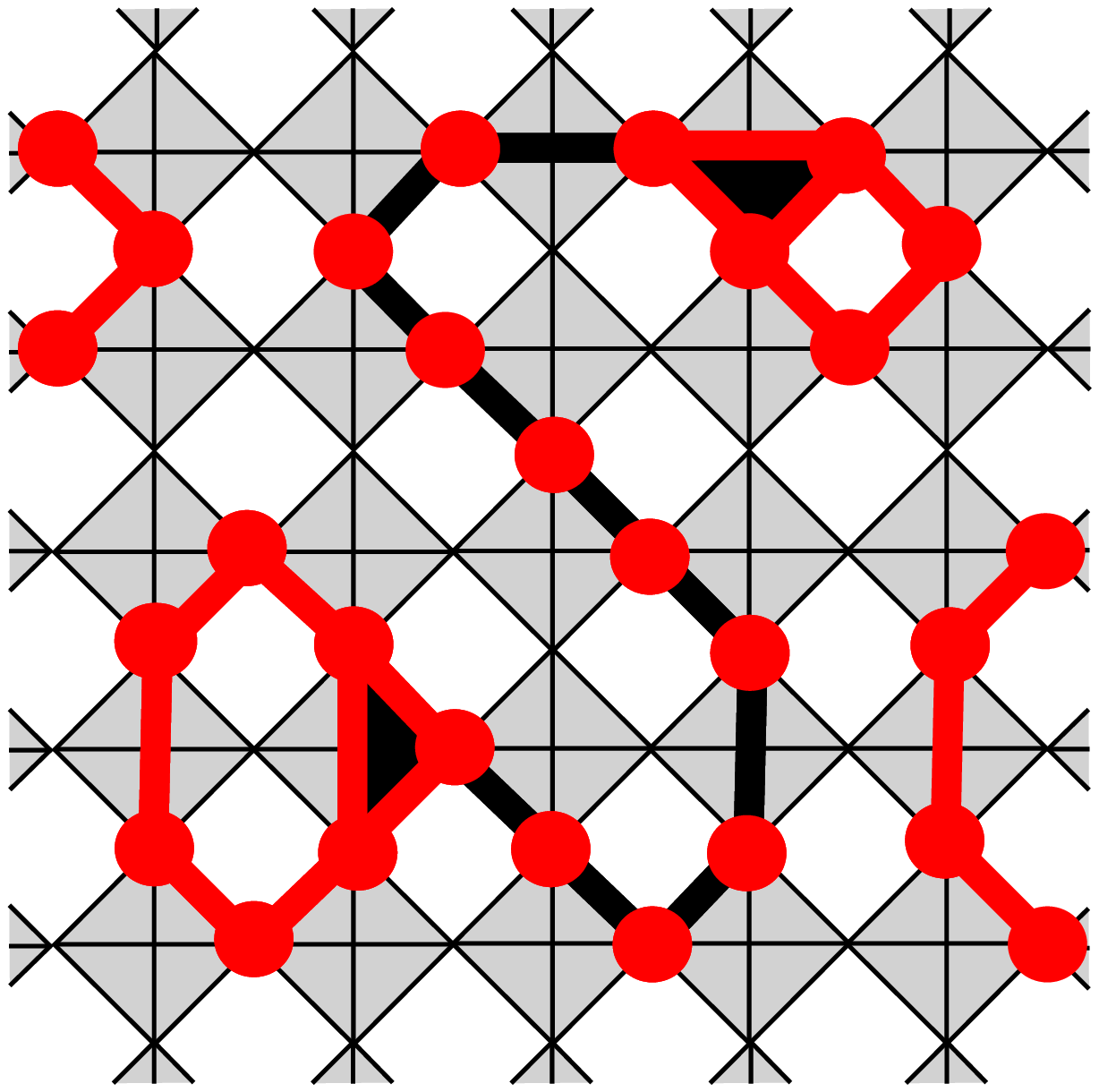}
\end{tabular}
\end{center}
\caption{(a) Adding one particle to the half-filled checkerboard lattice leads
to two defects (marked by black triangles) on adjacent crisscrossed squares. 
(b), (c) Two defects with charge $e/2$  can separate by particle hopping
without increase in repulsive energy. They are connected by a string consisting
of an odd number of occupied sites.
\label{cap:C2_added}}
\end{figure}

\subsection{Excitations with charge $\pm$e/2:}

\begin{figure}
\begin{center}
\begin{tabular}{llll}
(a)~\includegraphics[height=22mm,keepaspectratio]{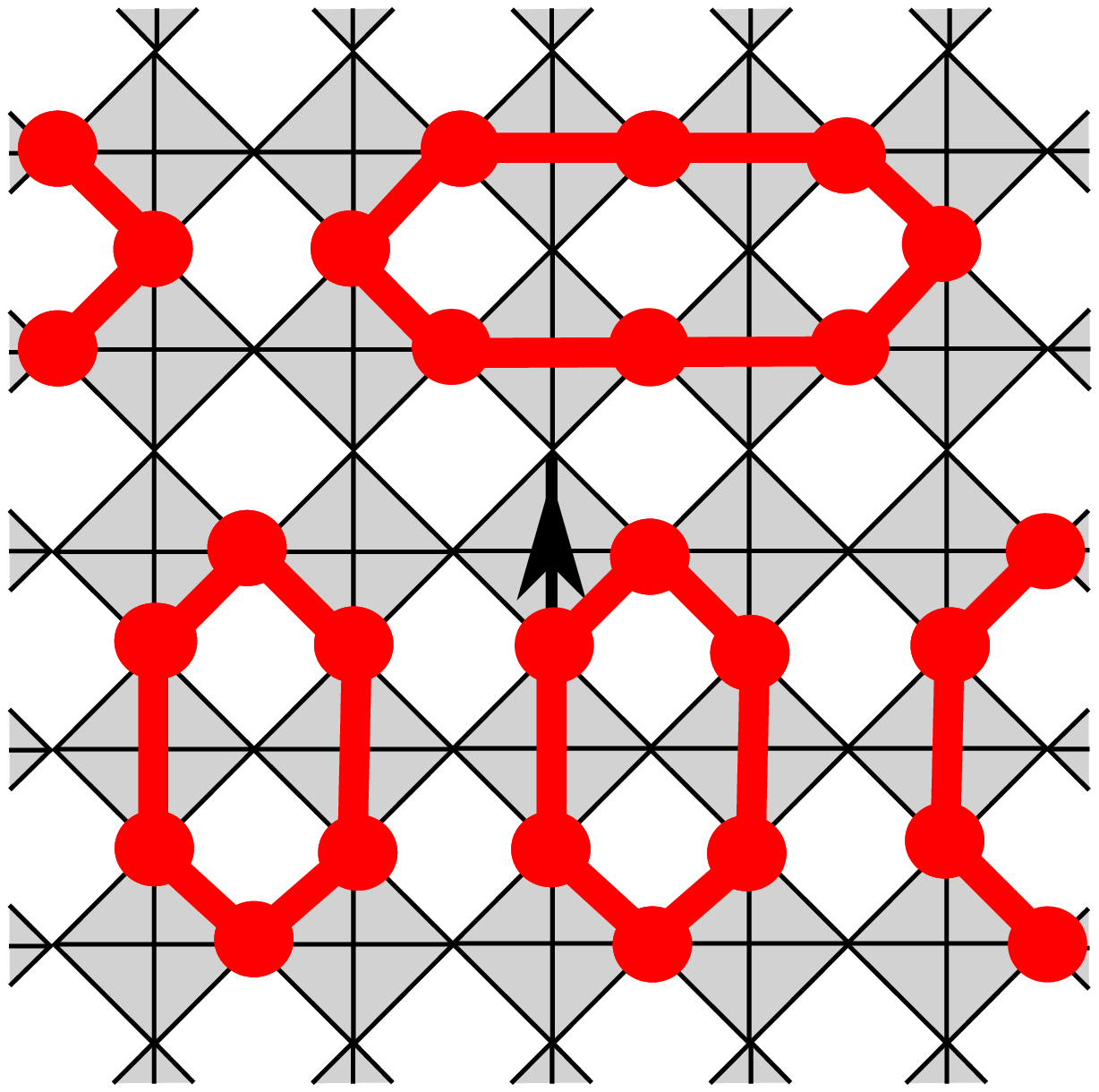}&
(b)~\includegraphics[height=22mm,keepaspectratio]{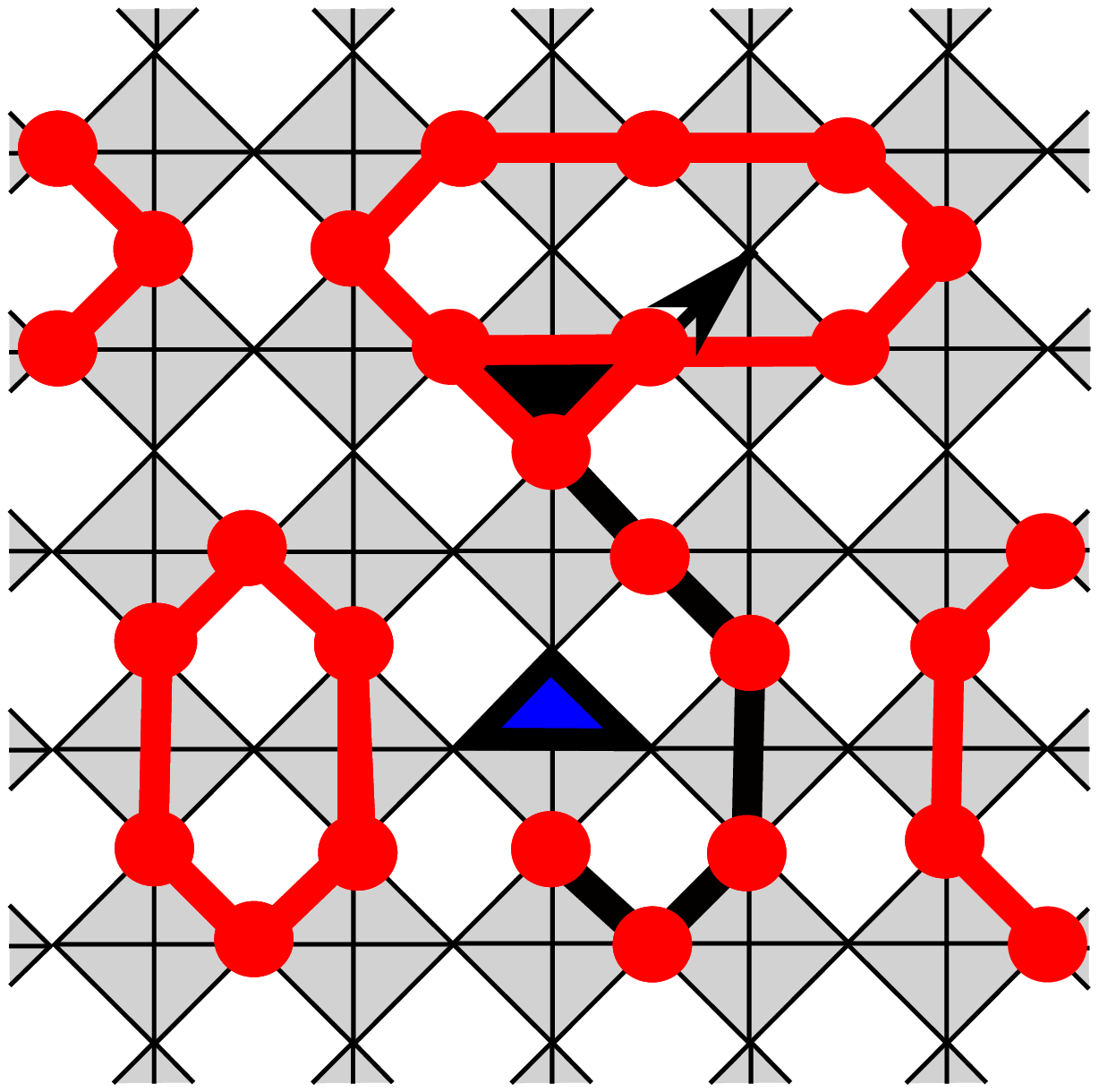}&
(c)~\includegraphics[height=22mm,keepaspectratio]{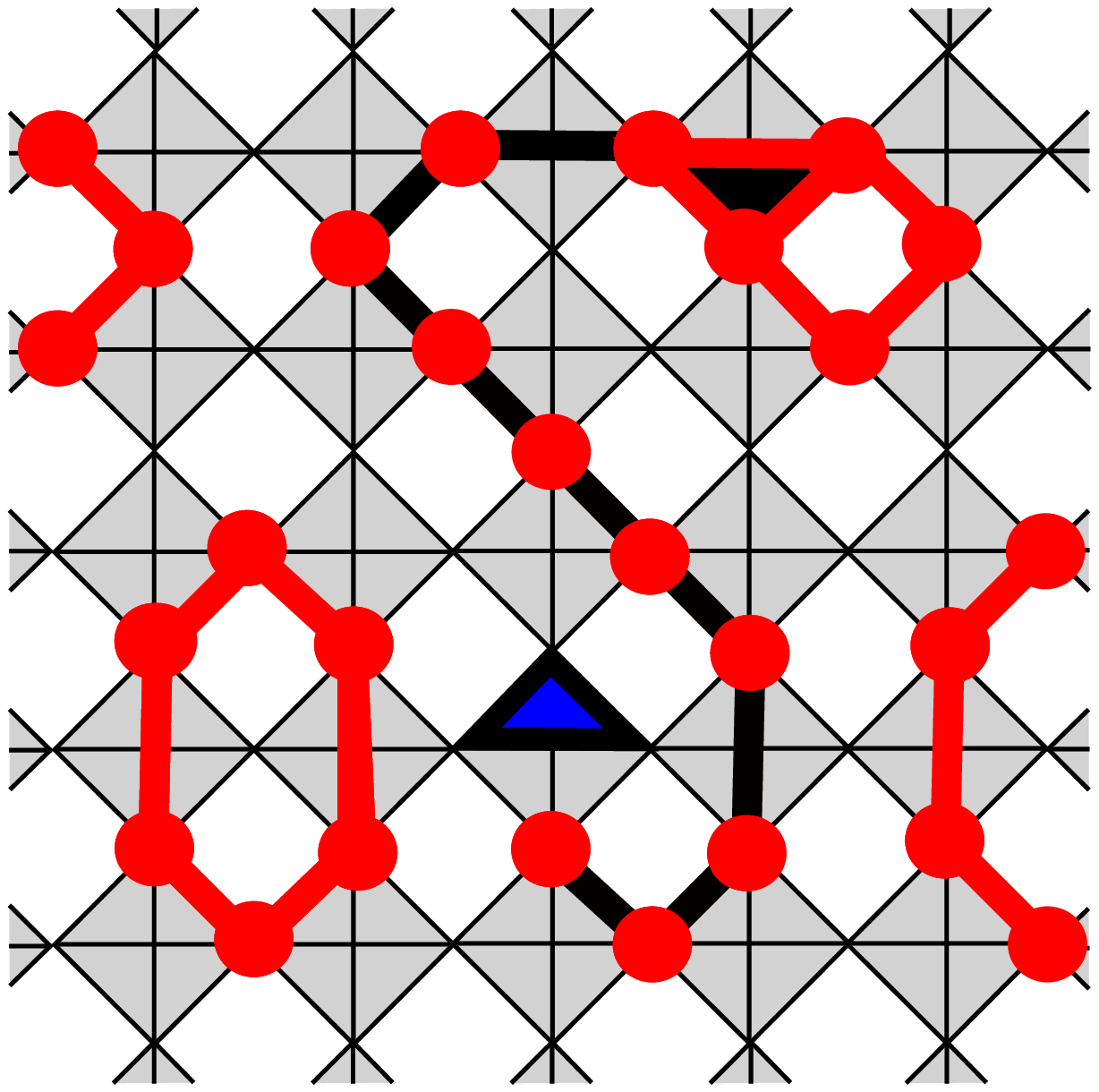}&
(d)~\includegraphics[height=22mm,keepaspectratio]{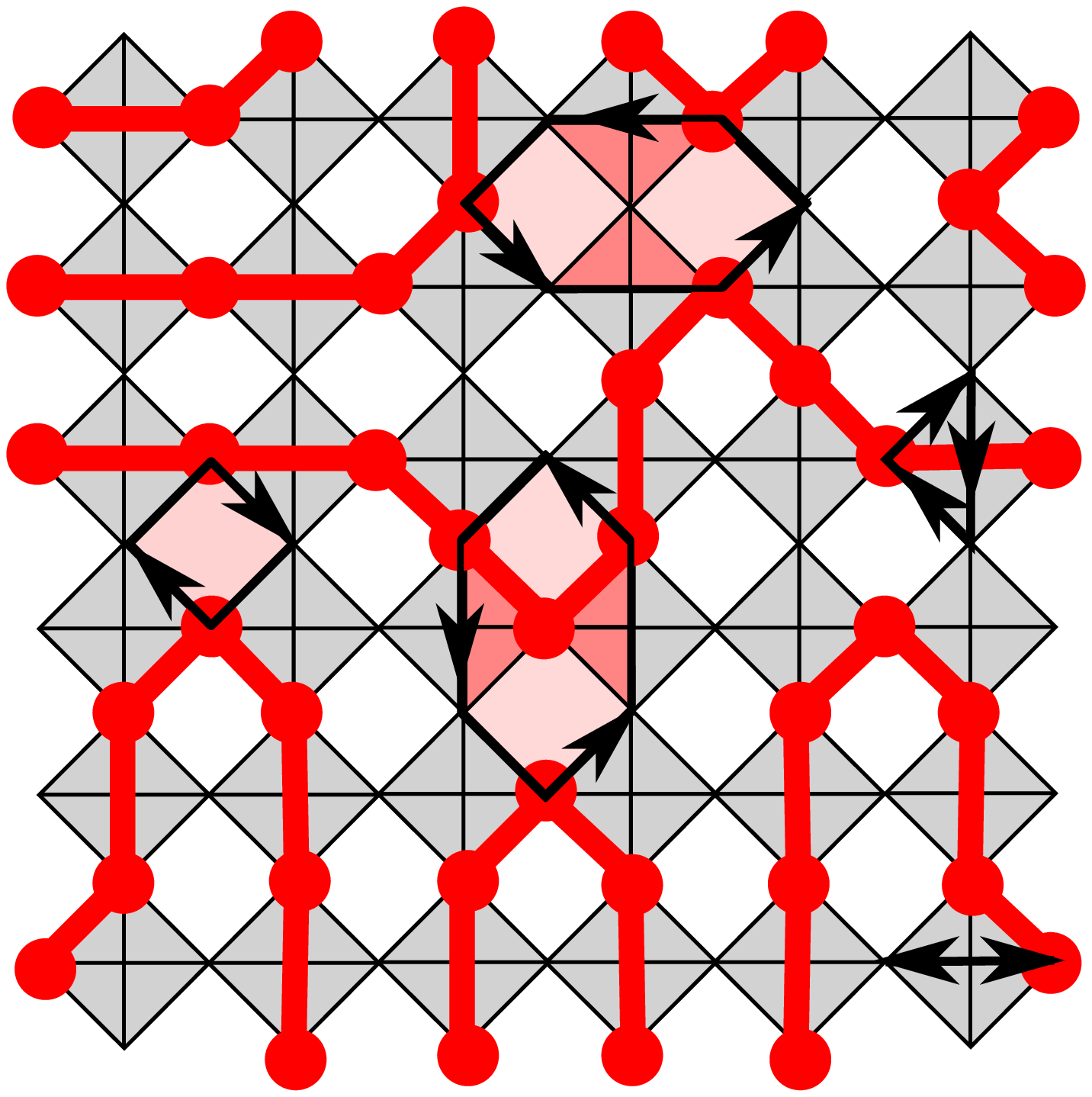}
\end{tabular}
\end{center}
\caption{(a)--(c) Hopping of a particle to a neighboring site: (a)~A
  fractionally charged particle and fractionally charged hole are
  generated. (b)--(c)~The two defects (marked by triangles) with charge $\pm
  e/2$ can separate without creating additional defects and are connected by a
  string consisting of an even number of particles.
\label{cap:C2_fluct}
(d) Example of an allowed configuration on a checkerboard lattice
at half filling with possible low-order hopping processes
[\onlinecite{Runge04}].  
\label{cap:C4_Checkerboard-lattice}}
\end{figure}

Assume that an electron is added to the otherwise half-filled lattice. In that
case two neighboring tetrahedra violate the tetrahedron rule because they
contain three particles each (see FIG. \ref{cap:C2_added}(a). The energy
required for adding the particle is 4$V$. Next assume that one particle is
hopping to a nearest neighbor site as indicated in
FIGs. \ref{cap:C2_added}(a,b). The new configuration has the same total
repulsive energy as the old one. This can continue as shown in
FIG. \ref{cap:C2_added}(c) and so on. One notices that the two tetrahedra with
three particles have separated without increase in repulsive energy. The charge
$e$ of the added particle has thus split into two fractions $e/2$. A similar
feature is observed if we add an energy $V$ to the ground-state energy. This
suffices to break a loop as indicated in
FIG. \ref{cap:C4_Checkerboard-lattice}(a,b). We notice a tetrahedron with three
particles and one with one particle only. 

Since a tetrahedron with three particles corresponds to a charge $e/2$, the one
with one particle only must correspond to a charge $-e/2$. It is an important
feature that the fractional charges are always connected by a string of
occupied sites, i.e., black lines in FIGs. \ref{cap:C2_added} and
\ref{cap:C4_Checkerboard-lattice}. In the case of two charges $e/2, e/2$ the
string contains always an odd number of sites while for a pair $e/2, -e/2$ the
number of sites on the connecting string is always even. When dynamics is added
to the system the fractional charges will separate because each of them will
gain kinetic energy. The energy of an added particle with momentum ${\bf
  k}$ is therefore 

\begin{equation}
E\left( {\bf k} \right) = 4V + \epsilon \left( {\bf k}_1 \right) + \epsilon
\left( {\bf k}_2 \right)
\label{eq:2}
\end{equation} 

\noindent with ${\bf k} = {\bf k}_1 + {\bf k}_2$. The form of $\epsilon ({\bf
  k})$ is yet unknown. Similarly the energy of a broken string or
  vacuum fluctuation is

\begin{equation}
\Delta E = V + \epsilon \left( {\bf k} \right) + \bar{\epsilon} \left( -{\bf k}
\right) 
\label{eq:3}
\end{equation}

\noindent where $\bar{\epsilon} ({\bf k})$ is the kinetic energy of the
fractional charge $-e/2$.

\begin{figure}
\begin{center}
\begin{tabular}{cc}
(a)\includegraphics[height=35mm,keepaspectratio]{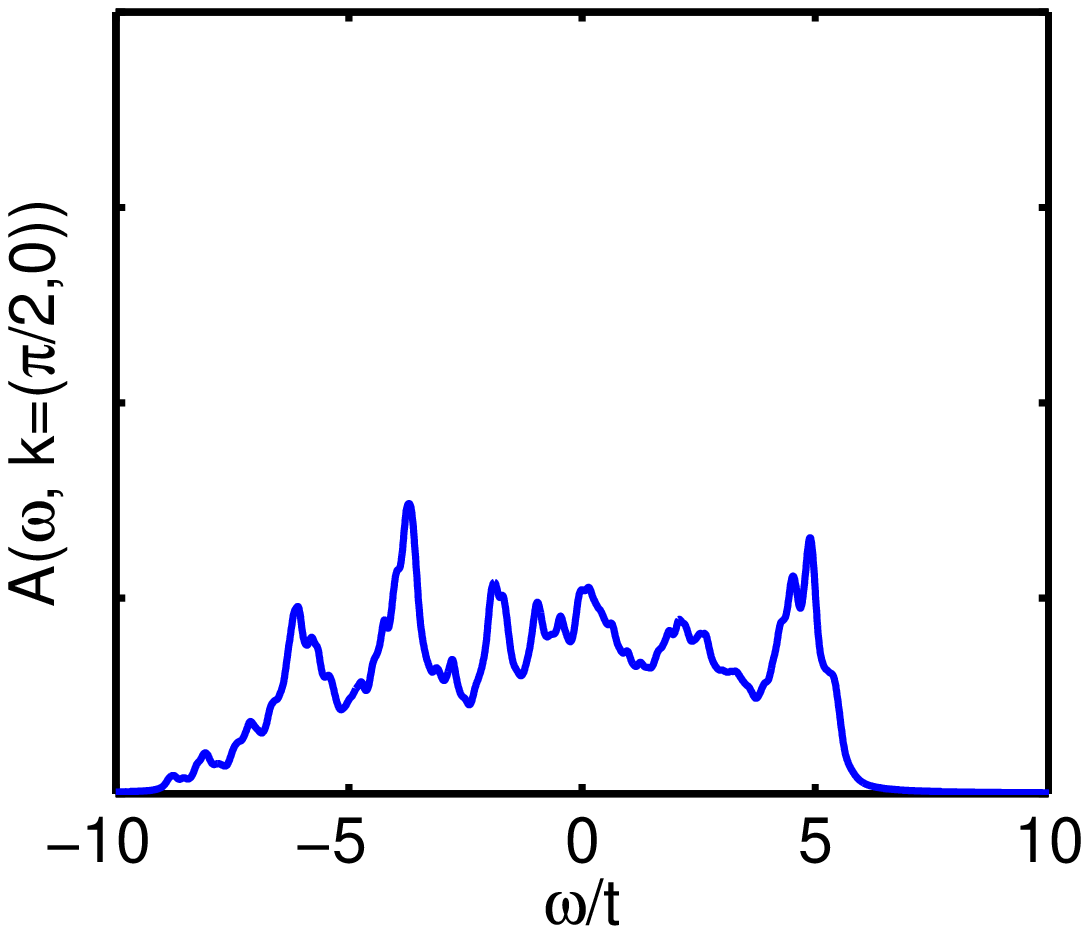}&
(b)\includegraphics[height=35mm,keepaspectratio]{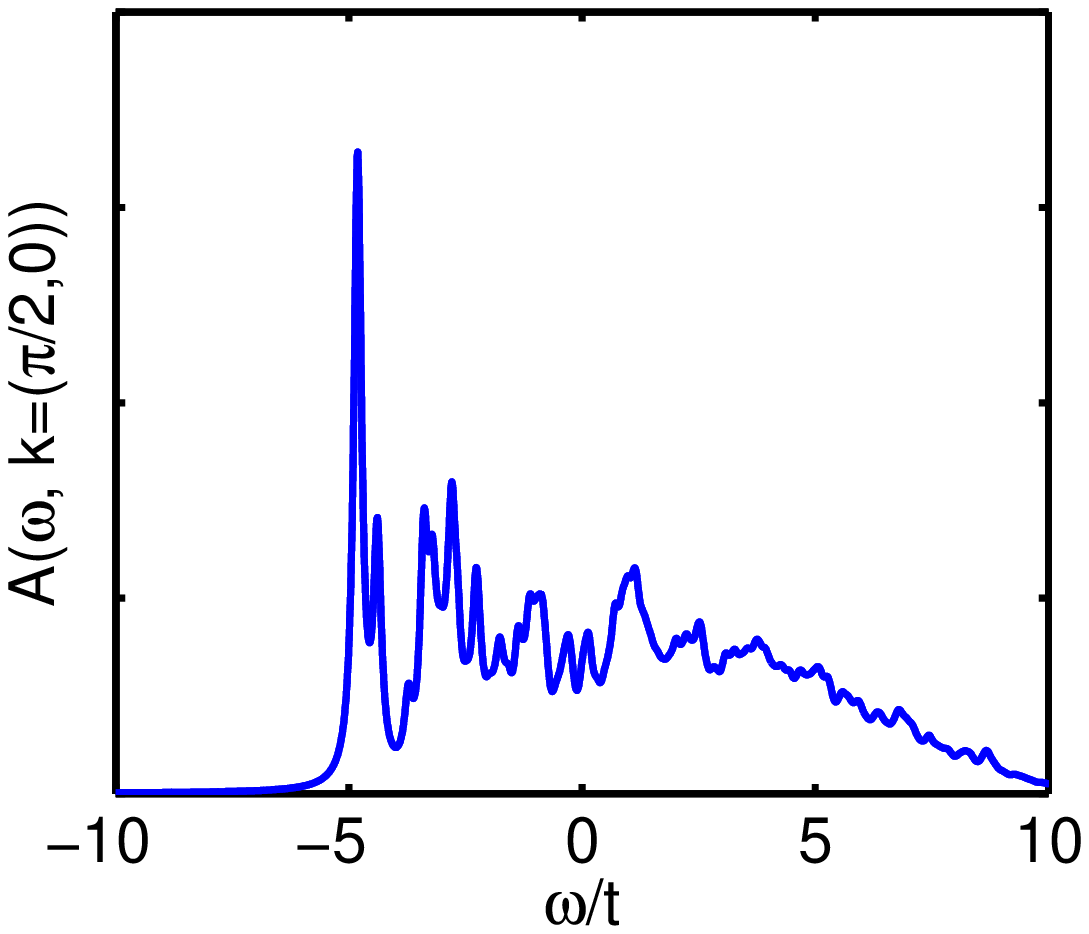}
\end{tabular}
\end{center}
\caption{(a) Spectral function $A(\mathbf{k}=\pi/2,\omega)$ for $V=25t$
  calculated for a $\sqrt{32}\times\sqrt{32}$ cluster for the effective
  $t$-$g$ Hamiltonian (\ref{eq:9}) with (a) $g=0.01$ and (b) $g=1$. A
  Lorentzian broadening $\eta=0.1t$ is used [\onlinecite{Pollmann06}].  
\label{cap:Spectral-density_gg}}
\end{figure}

Fractionally charged excitations are clearly outside the Landau concept of
Fermi liquids. Therefore the spectral function does not show a quasiparticle
peak like a Fermi liquid does (see
FIG. \ref{cap:Spectral-density_gg}(a). Instead it shows structures which are
partially due to finite size effects but have not yet been analyzed in
detail. The width of the spectral function is larger than for a particle which
is prevented from breaking up into two fractional charges $e/2$. A quasi-particle peak reappears if the ring exchange (see Section \ref{Sect:DynProc}) is so large that
confinement (see Section \ref{Sect:ConfCeconf}) becomes strong (see
FIG. \ref{cap:Spectral-density_gg}(b). A more detailed discussion of the
spectral function is given in Sect. \ref{Sect:ConfCeconf}. The remaining
question is whether the fractional charges are confined or deconfined.

\subsection{Side remark: inclusion of spin}

The above considerations referred to spinless fermions. When the spin is
included the question arises what happens to it when the charge fractionalizes
and falls apart into two pieces. To answer it we have to first generalize the
Hamiltonian (\ref{eq1}) to the form

\begin{equation}
H = -t\sum_{\langle i,j \rangle \sigma} \left( c_{i \sigma}^{\dag} c^{\vphantom{\dag}}_{j \sigma}
+ \mbox{H.c.} \right) + V \sum_{\langle i,j \rangle} n_{i} n_{j} + U \sum_i
n_{i \uparrow} n_{i \downarrow}~.  
\label{eq:4}
\end{equation}
\noindent  This is an extended Hubbard Hamiltonian with on-site repulsion $U$ and $n_i =
\sum\limits_\sigma c_{i \sigma}^{\dag} c_{i \sigma}^{\vphantom{\dag}}$. It is well known that in
the strong correlation limit it can be reduced to a $t-J$ Hamiltonian
[\onlinecite{auerbach94}] of the form

\begin{equation}
H_{t-J} = -t\sum_{\langle i,j \rangle \sigma} \left( \hat{c}_{i
  \sigma}^{\dag} \hat{c}^{\vphantom{\dag}}_{j \sigma} + \mbox{H.c.} \right) + V \sum_{\langle i,j
  \rangle} n_{i} n_{j} + \frac{J}{2} \sum_{\langle i,j \rangle} \left( {\bf S}_i {\bf
  S}_j - \frac{n_i n_j}{4} \right)  
\label{eq:5}
\end{equation}

\noindent where the operators

\begin{eqnarray}
\hat{c}_{i \sigma}^{\dag} & = & c_{i \sigma}^{\dag} \left( 1 - n_{i - \sigma}
\right)\nonumber \\
\hat{c}_{i \sigma} & = & c_{i \sigma} \left( 1 - n_{i - \sigma}
\right)   
\label{eq:6}
\end{eqnarray}

\noindent assure that sites are either empty or singly occupied, but never
doubly occupied while $J = 4t^2/U$ implies an antiferromagnetic
nearest-neighbor interaction. Consequently the strings which connect the
fractional charges represent Heisenberg chains with odd (in case of $e/2,
e/2$) or even (in case of $e/2, -e/2$) number of sites. But a chain with an odd
number of sites has a twofold degenerate ground state and therefore acts like
an effective spin $S = 1/2$. In distinction a chain with an even number of
sites has a singlet ground state corresponding  to $S = 0$. The spin of an
added particle is therefore distributed all over the chain and hence
delocalized over parts of the sample. This a rather unique physical
situation. It should be stressed that all of the above features are not
restricted to the checkerboard lattice but hold for the pyrochlore lattice as
well where they are less transparent though. A detailed study of the interplay between charge and spin degrees of freedom on the checkerboard lattice at fractional filling factors 1/4 and 1/8 can be found in Refs.~[\onlinecite{poilblanc2007a,poilblanc2007b,poilblanc2007c}].


\section{Dynamical Processes}

\label{Sect:DynProc}

The macroscopic degeneracy of the ground state of a half-filled checkerboard-
or pyrochlore lattice is lifted when dynamics is introduced, i.e., when $t$
with $|t| \ll V$ is taken into account [\onlinecite{Runge04}]. This lifting
takes place to order $t^3/V^2$, because contribution of order $t^2/V$ are equal
for all configurations. Note that ring hopping processes of order $t^2/V$
cancel for fermions. But in order $t^3/V^2$ ring hopping processes connect
different configurations with each other. Thus to that order the effective
Hamiltonian is of the form

\begin{equation}
H_{\rm eff} = \frac{12 t^3}{V^2} {\textstyle
  \sum\limits_{\{\smallhexh\}}} c_{j_6}^\dag c_{j_4}^\dag
c_{j_2}^\dag c_{j_5}^{\vphantom{\dag}} c_{j_3}^{\vphantom{\dag}} c_{j_1}^{\vphantom{\dag}}
\label{eq:7}
\end{equation}

\noindent and the sum is over all hexagons of the lattice. We rewrite this
expression in a more pictorial version as

\begin{eqnarray}
H_{\mathrm{eff}} & = & -g{\textstyle
  \sum\limits_{\{\smallhexh,\smallhexv\}}}\big(\,\big|\hexaf\big\rangle\big
\langle\hexbf\big|-\big|\hexafo\big\rangle\big\langle\hexbfo\big|+\mbox{H.c.}\,\big)\nonumber\\[2ex]  
& =: & -g{\textstyle \sum\limits_{\{\smallhexh,\smallhexv\}}} 
\Big( \big| B \big\rangle\big\langle \overline{B} \big| + \big| \overline{B}
  \rangle\langle {B} \big| -\big| A \big\rangle\big\langle \overline{A} \big| -
  \big| \overline{A} \big\rangle\big\langle {A} \big|\Big)  
\label{eq:C4_H_eff}
\end{eqnarray}

\noindent with coupling constant $g = 12 t^3/V^2>0$. Again the sum is taken
over all, i.e., vertically and horizontally oriented hexagons. The symbolic
form of $H_{\rm eff}$ is self-explanatory - occupied sites are indicated by a
dot, for examples see FIG. \ref{cap:C4_Checkerboard-lattice}(d). The minus signs
in Eq. (\ref{eq:C4_H_eff}) result from Fermi commutation rules associated with the
ring-hopping processes. They depend on the choice of how the multi-fermion states on the lattice are enumerated. The simple form of Eq. (\ref{eq:C4_H_eff}) is valid, e.g., for the enumeration used in \ref{Runge04} (along the diagonals)
In the case of doping $H_{\rm eff}$ is extended to

\begin{equation}
H_{t-g} = H_{\rm eff} - t \sum_{\langle ij \rangle} P \left( c_i^\dag c_j +
\mbox{H.c.} \right) P
\label{eq:9}
\end{equation}

\noindent where the projector $P$ projects onto the subspace of configurations
with the smallest possible violations of the tetrahedron rule compatible with
the number of doped particles. For one added particle the number of violations
is two. We call the extended Hamiltonian the $t-g$ model and consider $t$ and
$g$ as independent parameters. Thus we shall assume that the two parameters are
not necessarily restricted to $g \ll t$ as required by the strong correlation
limit. This is of advantage when numerical calculations are performed. The
Hamiltonian (\ref{eq:C4_H_eff}) and (\ref{eq:9}) does not include 8-site or
10-site hopping processes which become increasingly important for increasing
ratios $t/V$.


\section{Symmetries and conservation laws}

\label{Sect:SymConsLaw}

\begin{figure}
\begin{center}
\includegraphics[height=25mm,keepaspectratio]{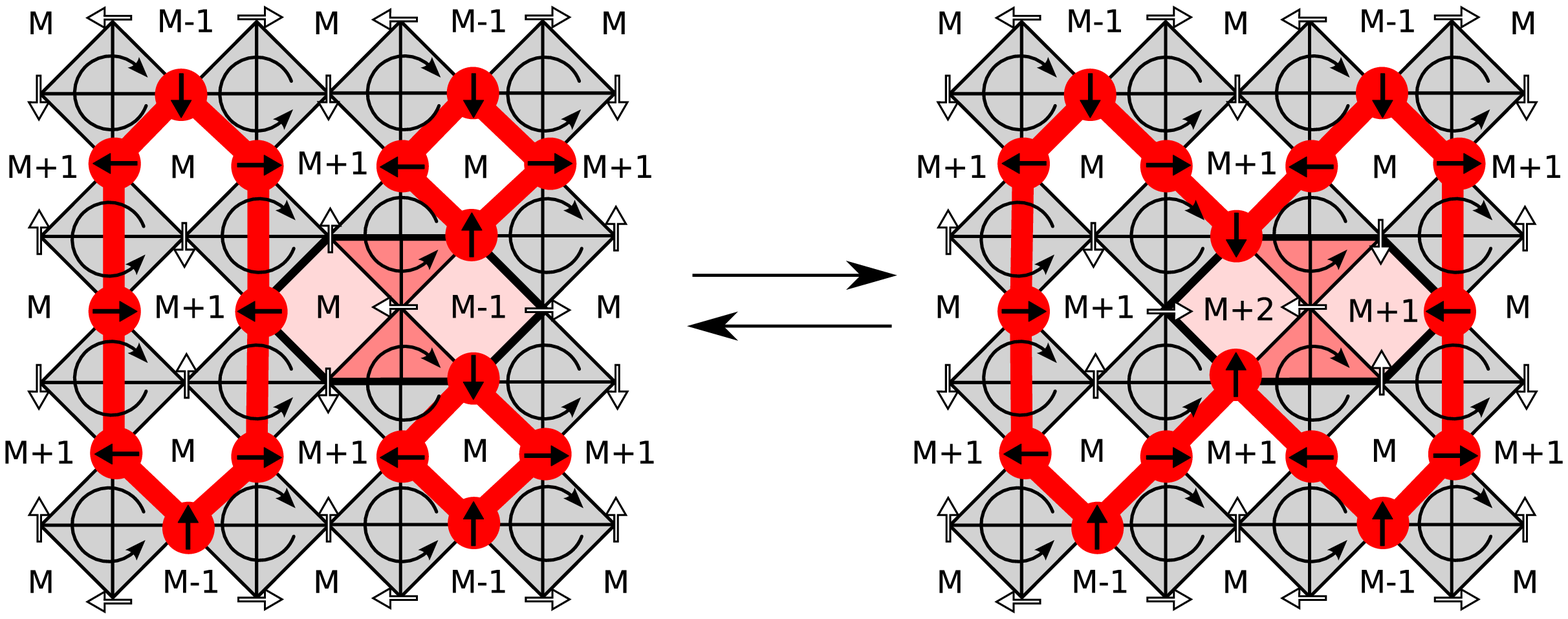}
\end{center}
\caption{Height representation for allowed configurations of a
  $\sqrt{32}\times\sqrt{32}$ checkerboard lattice with periodic boundary
  conditions at half filling. The height field $h$ (numbers in the non-crossed
  squares) is uniquely defined for a given configuration up to an additive
  constant $M$. The field $\mathbf{f}=\nabla h$ is indicated by small arrows on
  the lattice sites. Details of the mapping can be found in the text. The
  effect on the height fields of a ring-exchange process around a hexagon is
  shown explicitely [\onlinecite{Runge04}].
\label{cap:C4_height}}
\end{figure}

As pointed out before, $H_{\rm eff}$ lifts the macroscopic degeneracy of
allowed configurations, i.e., those which obey the tetrahedron rule. However,
the degeneracy is not totally lifted and a residual degeneracy remains. That
implies that the total manifold of allowed configurations is divided into
different subensembles within which all configurations are connected, but which
remain disconnected by $H_{\rm eff}$. For their identification it is useful to
determine the quantities which remain conserved by $H_{\rm eff}$. It is easy to
check that the particle numbers $N_1, ..., N_4$ on the four sublattices into
which the checkerboard (or pyrochlore) lattice can be divided remain invariant
under the operation of $H_{\rm eff}$. Furthermore, any allowed configuration of
the checkerboard lattice can be uniquely represented by a vector field ${\bf
  f}$ with a vanishing discretized lattice version of curl ${\bf f}$. This
vector field is obtained for the bipartite lattice by assigning an
alternating direction of orientation (i.e., clockwise and counter clockwise) to
the tetrahedra, i.e., crisscrossed squares. Each occupied site is associated
with an unit vector in the direction of orientation and each empty site with
one in opposite direction. This results in a mutual cancellation of the ${\bf
  f}$ vectors on a tetrahedron with two occupied and two empty sites or more
generally when a closed loop is formed. From curl ${\bf f} = 0$ it follows that
the vector field can be represented by ${\bf f}$ = grad $h$ where the scalar
field $h$ defines a hight field up to an arbitrary constant. The height at the
upper (right) and lower (left) boundary of a finite lattice of $N_x \cdot N_y$
crisscrossed squares can differ only by an integer $-N_{y(x)} \le \kappa_{y(x)}
\le N_{y(x)}$ when periodic boundary conditions are applied. It is the same for
all columns (rows) and defines two topological quantum numbers ($\kappa_x,
\kappa_y$). They are invariant under hexagon hopping processes. As seen from
FIG. \ref{cap:C4_height} the application of $H_{\rm eff}$ merely changes the
local height of two neighboring plain squares by $\pm$2. The doublet
$(\kappa_x/N_x, \kappa_y/N_y)$ specifies the global slope of the height field.

States with $(\kappa_x, \kappa_y) \neq (0, 0)$ have a broken lattice symmetry
and are charge ordered. For example, when $\kappa_x/N_x > 0$ the charge is
modulated along a diagonal stripe. Note that violation of the equality $N_1 =
N_2 = N_3 = N_4$ also implies a modulation of the charge density.

\begin{figure}
\begin{center}
\includegraphics[width=68mm]{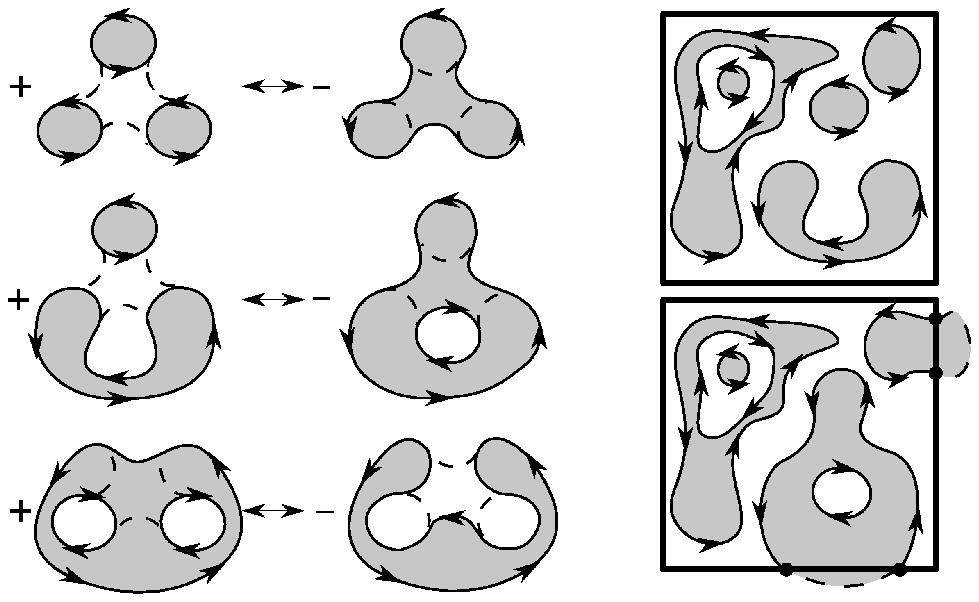}\\
\hspace*{2.5cm}(a)\hspace*{41mm}(c) \\[-2.9cm]
\hspace*{2.5cm}\phantom{(a)}\hspace*{41mm}(b) \\[2.5cm]
\end{center}
\caption{(a) Changes in loop topologies by ring hopping processes of Type A
  (see Eq. (\ref{eq:C4_H_eff})). (b), (c) Representations of two
  configurations by fully-packed directed loops [\onlinecite{Pollmann06c}]. 
\label{cap:C4_loops}}
\end{figure}

The different signs in $H_{\rm eff}$ poses a problem when quantum Monte
Carlo simulations are applied to the system. Therefore it is of considerable
help that in important cases this relative sign can be removed by a proper
gauge transformation. The system can then be transformed to a bosonic
one. Without going into details which are found in
Ref. [\onlinecite{Pollmann06c}] we describe merely the way this is achieved. We
consider allowed configurations with fixed boundary conditions with an even
number of fermions on the four boundaries. Open loops are closed as indicated
in FIG. \ref{cap:C4_loops}(c) . A given loop covering $\mathscr{L}$ of the plane
is processed as follows. We color the background white and alternate the color
inside a loop, either from white to dark or reverse. We attach a direction to
each loop so that the white colored regime is always to the right of the
loop (see FIG. \ref{cap:C4_loops}(b,c)). Next we count the total number $r$ of
clockwise and $l$ of counter-clockwise loops. The three types of topological
changes caused by hexagon flipping processes with empty center site  are shown
in FIG. \ref{cap:C4_loops}(a). Note that processes with an occupied center site
merely deform loops but do not change their topology. The sign change of the
processes with an empty center of the hexagon can be cured by transforming the
configuration $| \mathscr{L} \rangle$ to

\begin{equation}
\mid \mathscr{L} \rangle \rightarrow i^{l(\mathscr{L})} (-i)^{r(\mathscr{L})}
\mid \mathscr{L} \rangle ~~~. 
\label{eq:10}
\end{equation}

The above considerations need not to apply to periodic boundary conditions
where only subensembles with even winding numbers allow for the coloring
prescription given above. Even then it may happen that the coloring is reversed
by application of $H_{\rm eff}$ while the same loop configuration is
recovered. However, from numerical analysis it is found that for tori which
preserve the bipartiteness of the lattice the transformation applies indeed to
the states of lowest energy. In FIG. \ref{cap:C4_Energies-gmu}(a) we compare the
energies of a system consisting of 72 sites in which the Fermi sign is taken
into account and when it is gauged away, i.e., when the system is treated as a
bosonic one. The ground-state energy, the first excited states in the
$(\kappa_x, \kappa_y) = (0, 0)$ sector and the weights of the different
configurations are the same in both cases. When $(\kappa_x, \kappa_y) \neq (0,
0)$ we find that in some subensembles the ground-state energies are higher for
fermions than they are for bosons, i.e., when the fermionic sign is removed by
a gauge transformation. All the considerations are limited though to hexagon
ring hopping processes and do not apply to higher-order processes.

\begin{figure}
\begin{center}
\includegraphics[height=40mm]{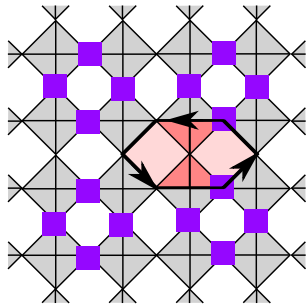}\\
\end{center}
\caption{Ring hopping changes the number of particles by two on sites marked by
  blue squares by two.
\label{cap:C4_The-four-sublattices}}
\end{figure}

It should be also mentioned that the sign of $g$ is irrelevant. By multiplying
all configurations by a factor $i^{\sigma_p}$ where $\sigma_p$ is the number of
fermions on the sublattice in FIG. \ref{cap:C4_The-four-sublattices}, the
sign of all matrix elements of $H_{\rm eff}$ is changed. Invariance under this
gauge transformation demonstrates a global $g \leftrightarrow -g$ symmetry of the spectrum. 

\begin{figure}
\begin{center}
(a)~\includegraphics[height=42mm]{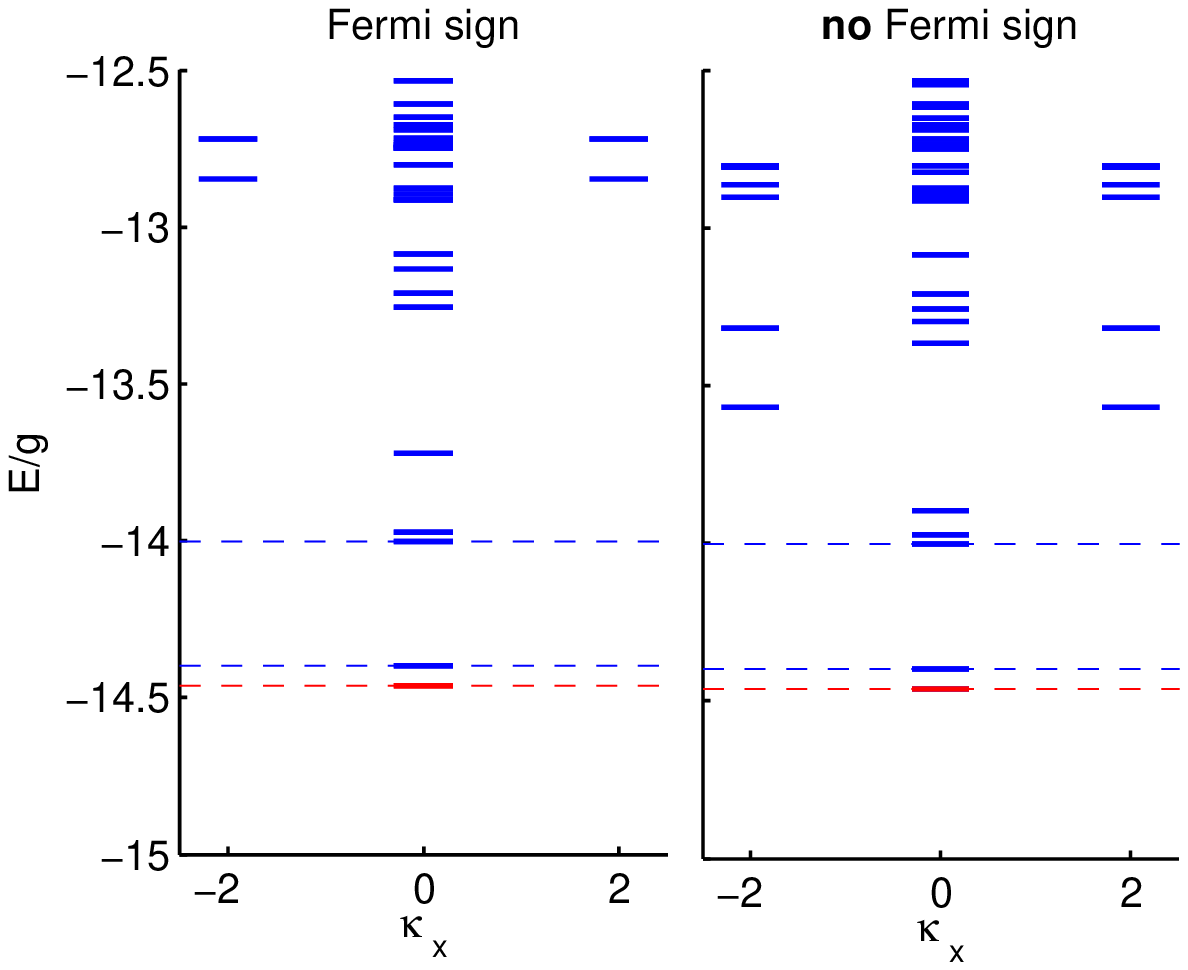}\\
(b)~\includegraphics[height=42mm]{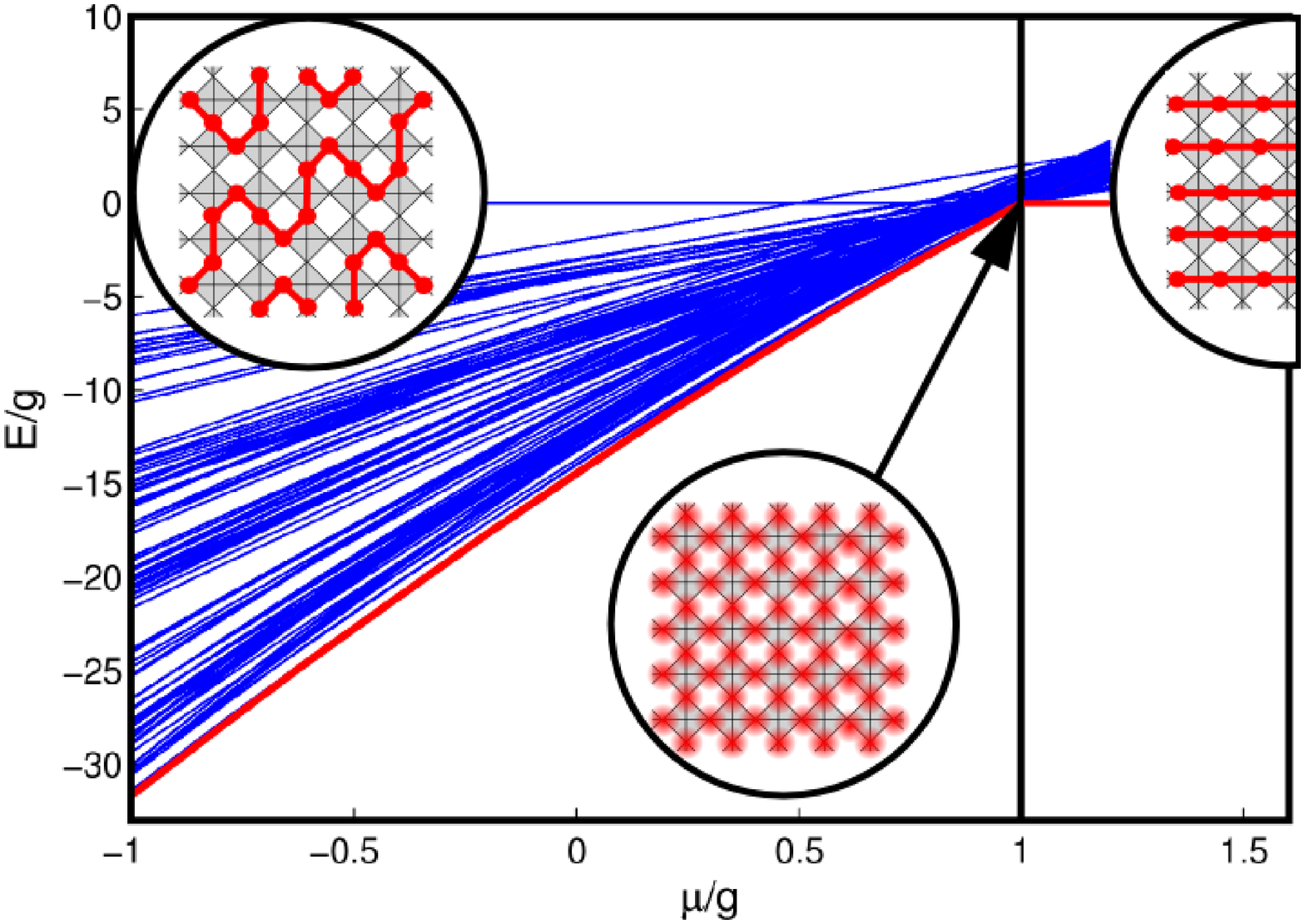}
\end{center}
\caption{(a) Ground-state energy and energies of the lowest excited states of
  the effective Hamiltonian $H_{\rm eff}$ in subspaces with different global
  slopes $(\kappa_x, \kappa_y) = (\kappa_x, 0)$ for a 72-site cluster. Right
  side: Same system, but assuming same signs for all matrix elements (''bosonic
  calculation''). (b) Energies of the ground state and lowest excited states
  for all subensembles of a 72-site half-filled checkerboard cluster for
  different values of $\mu$ of the $g$-$\mu$ Hamiltonian. Level crossing of
  ground states occurs only at $\mu=g$. The insets indicate different phases:
  Maximal flippable plus fluctuations for $\mu<g$, a critical point $\mu=g$
  where the ground state is an equally weighted superposition of all
  configurations, and frozen configurations as ground states for $\mu>g$,
  [\onlinecite{Pollmann06c}].} 
\label{cap:C4_Energies-gmu}
\end{figure}


\section{Confinement vs. deconfinement}

\label{Sect:ConfCeconf}

The question of confinement of fractional charges is closely related to the
form of the ground state of the half-filled lattice. For example, when the
ground state is charge ordered then one expects the fractional charges of a
doped system to be confined, because a separation of them results in disorder
of the ground state. Therefore we discuss first the character of the
ground-state wavefunction. The Hamiltonian is given by Eq. (\ref{eq:C4_H_eff})
but we want to supplement it following Rokhsar and Kivelson
[\onlinecite{Rokhsar88}] by an extra term which counts the number of flippable
hexagons. Depending on its sign, configurations with such hexagons are either
favoured or partially suppressed. The Hamiltonian which we consider reads
therefore 

\begin{eqnarray}
H_{\mathrm{g}\mu} & = & H_{\mathrm{eff}} + \mu{\textstyle
  \sum_{\{\smallhexh,\smallhexv\}}}
\big(|\hexafg\rangle\langle\hexafg|+|\hexbfg\rangle\langle\hexbfg|\big) 
\label{eq:C4_H_gmu}
\end{eqnarray}

\noindent where the sum is over all flippable hexagons with occupied or empty
center. Two limiting cases are particularly simple:\\

(i)~~~ \begin{minipage}[t]{15.0cm}
$\mu \rightarrow + \infty$: all configurations without flippable
  hexagons (frozen configurations) are ground states with energy $E = 0$.\\
\end{minipage}

(ii)~~ \begin{minipage}[t]{15.0cm}
$\mu \rightarrow - \infty$: configurations with maximal number of
  flippable hexagons $N_{fl}$ become ground states. Numerical Monte Carlo
  calculations for checkerboard lattices up to 1000 sites yield configurations
  with a ''squiggle'' type structure [\onlinecite{Pollmann06b}] or slight
  variations of it as ground states. The results suggest that in the
  thermodynamic limit they are all in the $(\kappa_x, \kappa_y) = (0, 0)$
  subspace. Due to the large unit cell of the squiggle configuration and the
  conservation laws the number of different ground states is ten.\\
\end{minipage}

(iii)~ \begin{minipage}[t]{15.0cm}
$\mu = g>0$: This is a particularly interesting case because it is
  exactly solvable [\onlinecite{Rokhsar88}]. The ground states have energy $E =
  0$ and some of them are liquid like. By using the gauge transformation
  (\ref{eq:10}) we change the sign of the second term in (\ref{eq:C4_H_eff})
  and rewrite $H_{g=\mu}$ in the form\\ 
\end{minipage}

\begin{eqnarray}
H_{\mathrm{g}=\mu} & = & g \sum_{\{\smallhexh,\smallhexv\}}
\big[\big(|\!\hexafg\!\rangle-|\hexbfg\!\rangle\big)\times\big(\langle\!\hexafg|-\langle\hexbfg\!|\big)\big] 
\label{eq:C4_H_RK_II}
\end{eqnarray}

\hspace{0.7cm}\begin{minipage}[t]{15.0cm}
which is a sum over projectors. Therefore all eigenvalues must be
non-negative. For each subensemble $l$ the ground-state wavefunction is then given by an
equally weighted superposition of all connected configurations
$|c_i^{(l)}\rangle$, i.e.,
\end{minipage}

\begin{equation}
\mid \psi_0^{(l)} \rangle = A \sum_i \mid c_i^{(l)} \rangle
\label{eq:13}
\end{equation}

\hspace{0.7cm}\begin{minipage}[t]{15.0cm}
where $A$ is a normalization prefactor. Using the Perron and Frobenius  Theorem (see below), it follows that these are the unique ground states in each subensemble. The coherent superpositions
are resonating Valence Bond (RVB) states of the form originally discussed in
[\onlinecite{anderson73,fazekas74}]. We want to draw
attention that a RVB state is obtained only when a gauge transformation of the
form (\ref{eq:10}) exists which ensures that all off-diagonal matrix
elements have the same sign. Otherwise the energy is most likely larger than
zero.\\
\end{minipage}

Next we explore the phase diagram as function of $\mu/g$ by exact
diagonalization of $H_{g \mu}$ for clusters up to 72 sites. The details are
found in Ref. [\onlinecite{Pollmann06b}] so that we state here merely the
results:\\ 

(i)~~~ \begin{minipage}[t]{15.0cm}
For $\mu > g$ we find the same features as for ${\mu \rightarrow
  \infty}$, i.e., all frozen configurations are ground states.\\
\end{minipage}

(ii)~~ \begin{minipage}[t]{15.0cm}
For $\mu < g$ we find two ground states in subensembles with $(\kappa_x,
  \kappa_y)$ = (0,0). They are ($N_1, ... , N_4$) = (6, 6, 12, 12) and (12, 12,
  6, 6) respectively. They consist of superposition of configurations with
  maximal number of flippable hexagons and of fluctuations around these
  configurations. We expect that in the thermodynamic limit the 10-fold
  degenerate squiggle phase instead of a two-fold degenerate ground state is
  recovered. The average weight of configurations with maximal number of
  flippable hexagons decreases as $\mu$ increases until at $\mu = g$ all
  configurations of a subensemble have the same weight.\\
\end{minipage}

\noindent The energy of the ground state and of the lowest excited states as
function of $\mu/g$ is shown in FIG. \ref{cap:C4_Energies-gmu} for a cluster of
72 sites. 

\begin{figure}
\begin{center}
\begin{tabular}{lll}
(a)~\includegraphics[height=30mm]{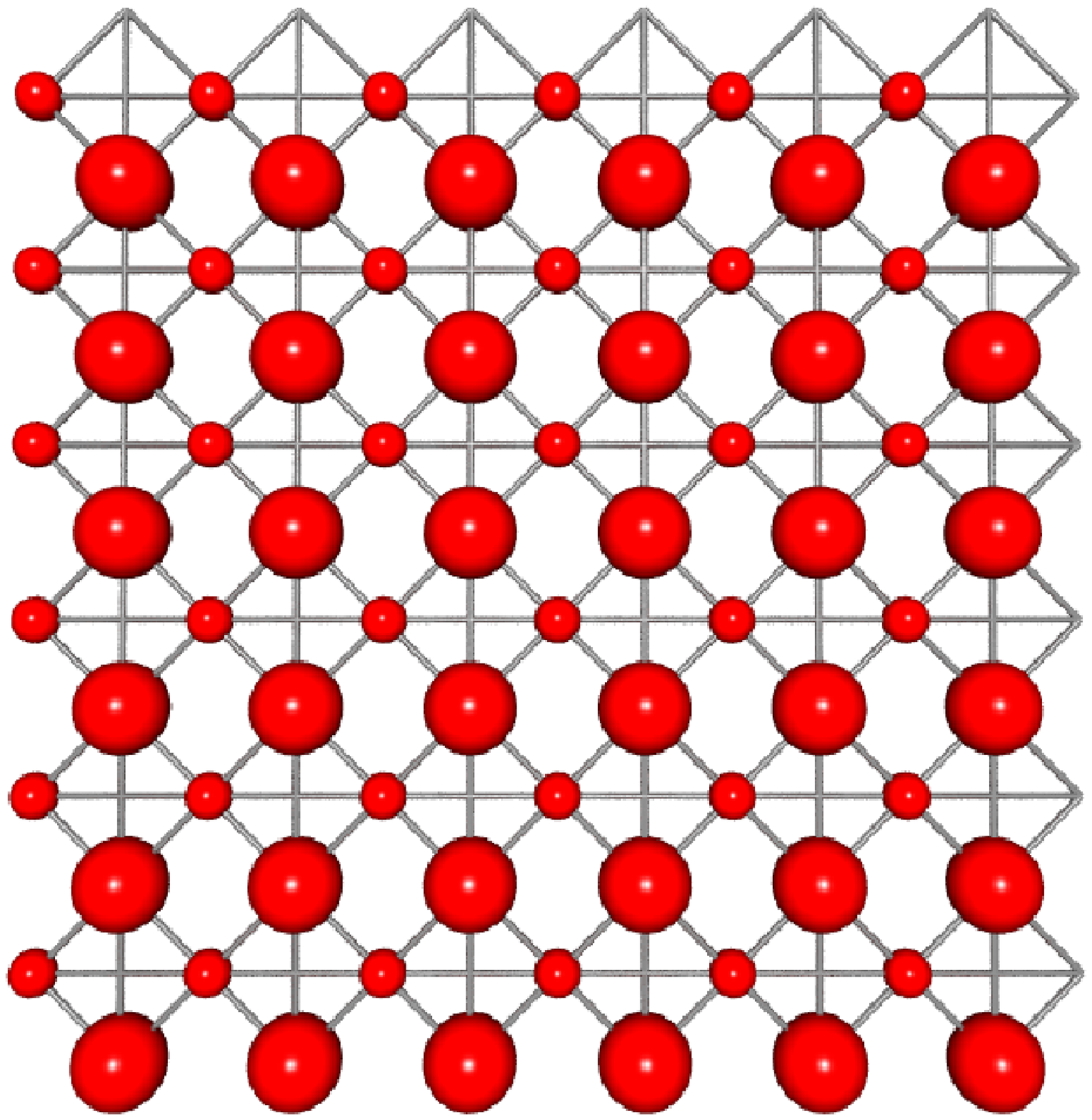}&
(b)~\includegraphics[height=30mm]{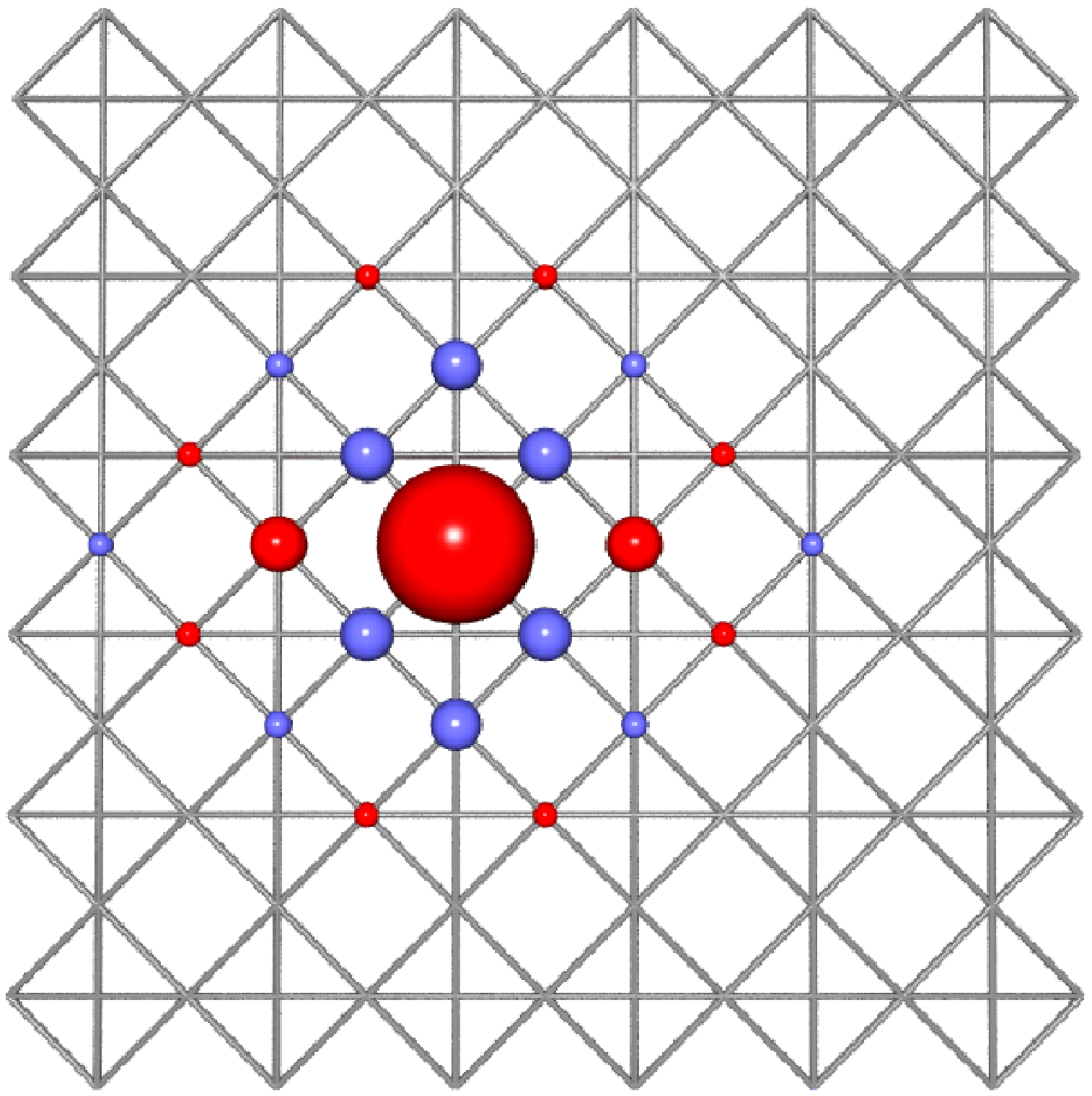}&
(c)~\includegraphics[height=30mm]{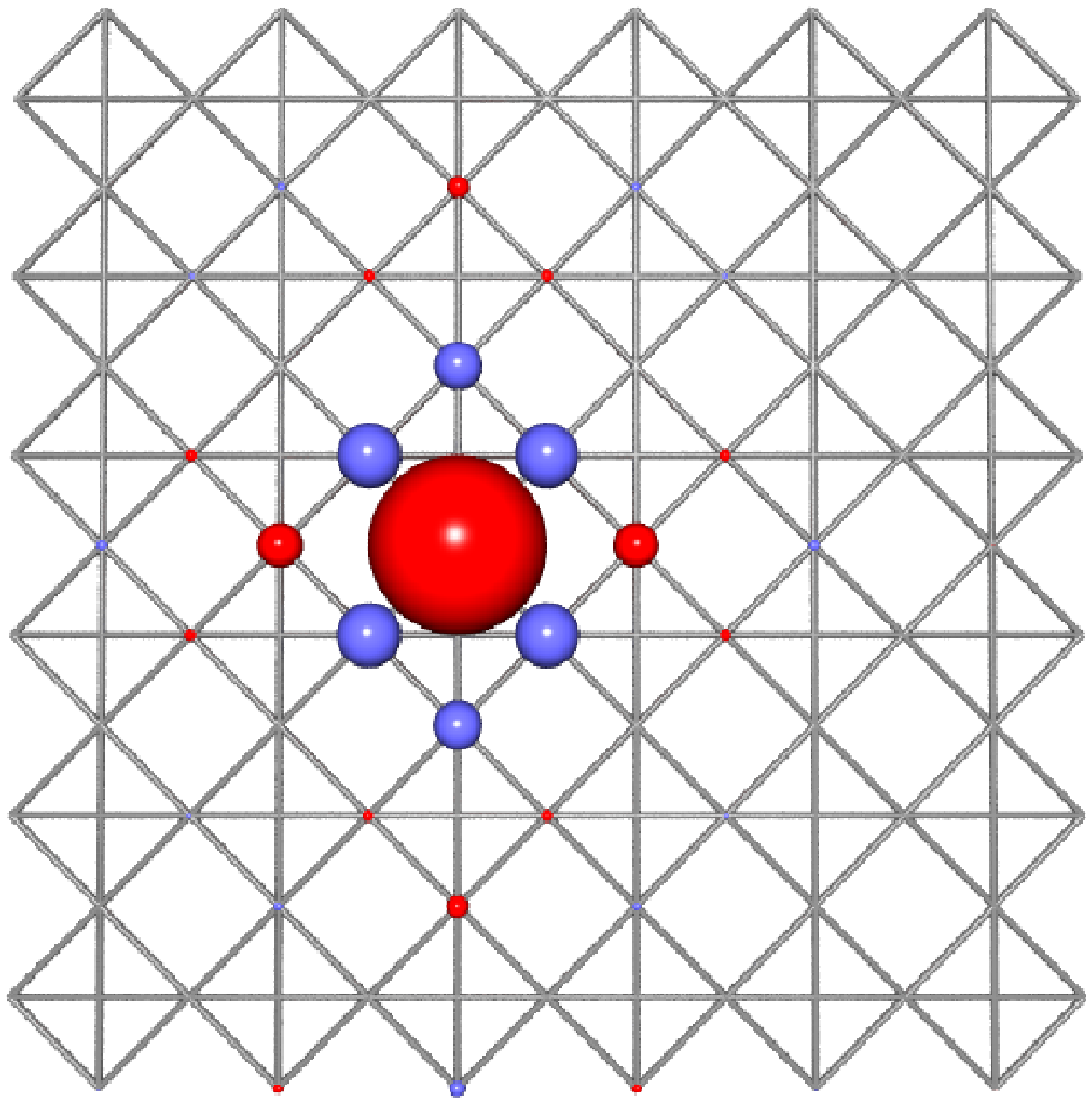}
\end{tabular}
\end{center}
\caption{Half filling: (a) Charge density distribution for one of the two
ground states. (b) Corresponding density-density correlation function
$C_{i_{0}i}^{(l)}$. The site $i_{0}$ with average density 2/3 shows
up as the largest dot in the panels (b) and (c). The radius of the dots
is proportional to the absolute value. Red or blue color represents
a positive and negative value, respectively. (c) Classical density-density
correlation function. 
\label{cap:C4_Panels_nc_HF}}
\end{figure}

It is interesting to calculate the density-density correlation function for a
72-site cluster at $\mu = 0$ when one of the two degenerate ground states is
used for $\psi_0$, i.e.,

\begin{equation}
C_{i0} = \langle \psi_0 \mid n_i n_0 \mid \psi_0 \rangle - \langle \psi_0 \mid
n_i \mid \psi_0 \rangle \langle \psi_0 \mid n_0 \mid \psi_0\rangle~~~.
\label{eq:14}
\end{equation}

\noindent Here the subscript 0 denotes a site near the center of the
cluster. The charge order for the state $N_1, ... , N_4$ = (6, 6, 12, 12)
yields stripes of average occupation 1/3 and 2/3 (see
FIG. \ref{cap:C4_Panels_nc_HF}(a). The result for $C_{i0}$ is shown in
FIG. \ref{cap:C4_Panels_nc_HF}(b). It differs only slightly from the
corresponding classical correlation function plotted in
FIG. \ref{cap:C4_Panels_nc_HF}(c) which is obtained by setting $t = 0$ and
summing over all degenerate configurations. 

\begin{figure}
\begin{center}
\begin{tabular}{ccc}
(a)\includegraphics[width=27mm]{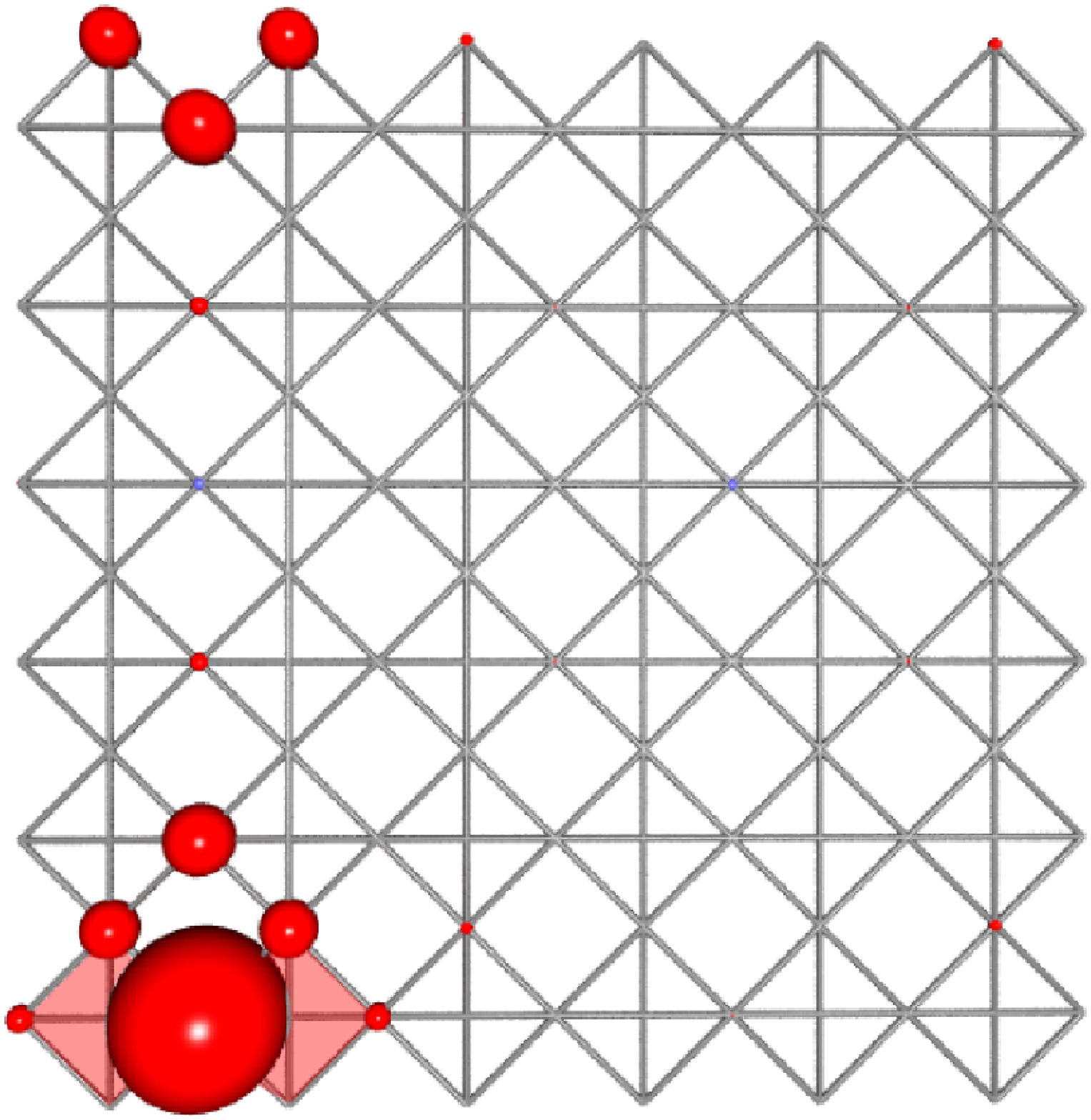}&
(b)\includegraphics[width=27mm]{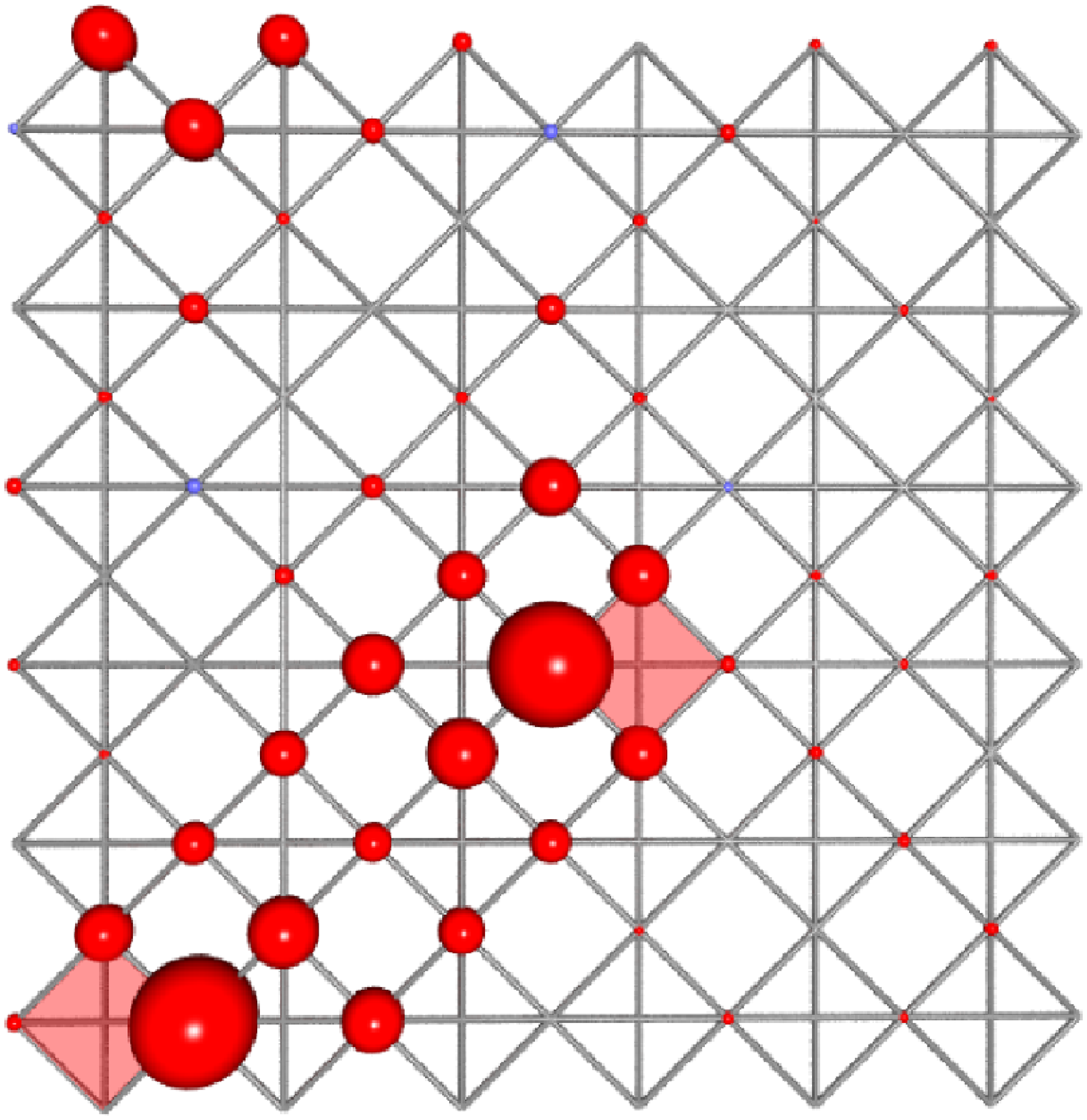}&
(c)\includegraphics[width=27mm]{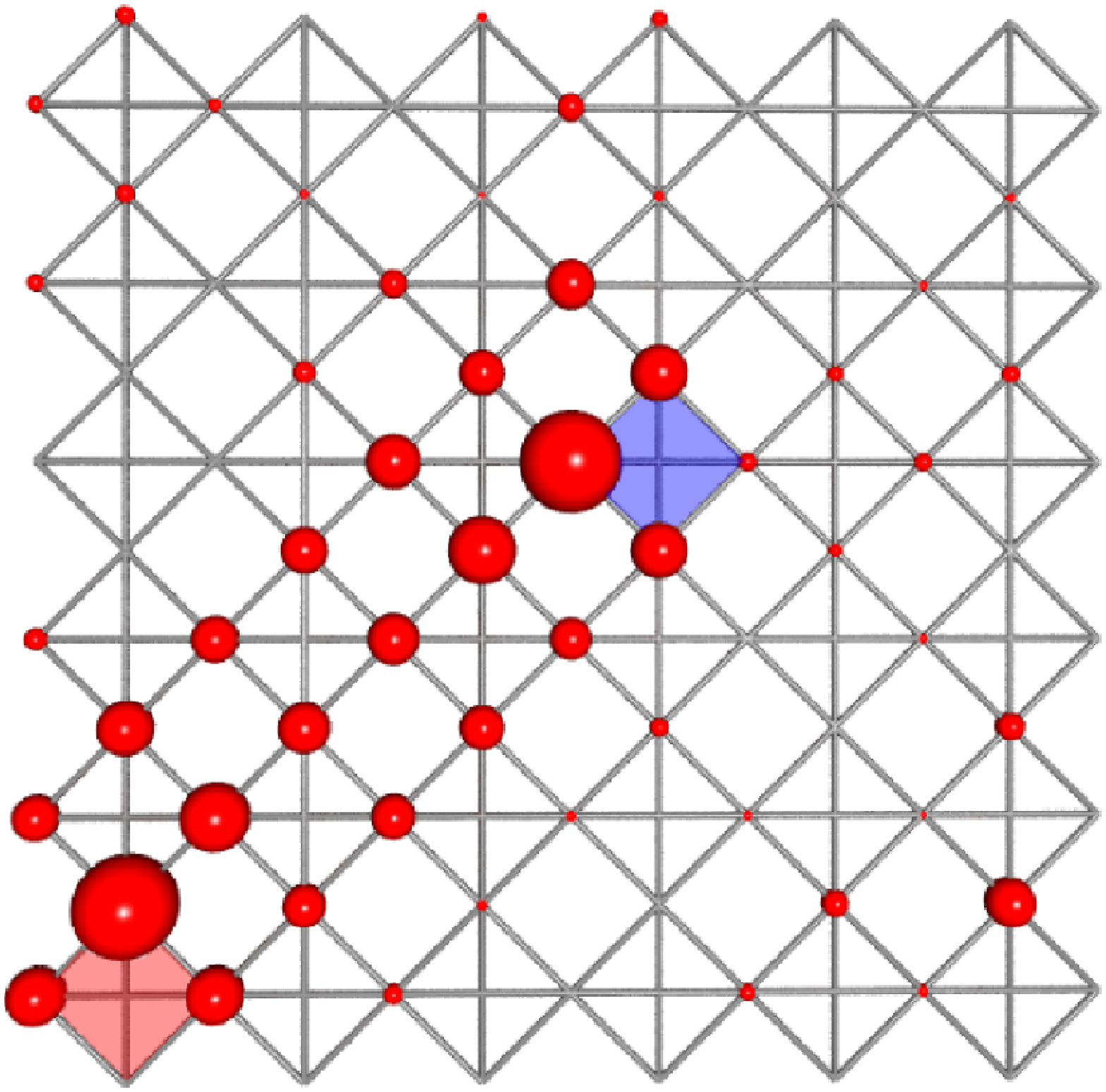}\\ 
(d)\includegraphics[width=27mm]{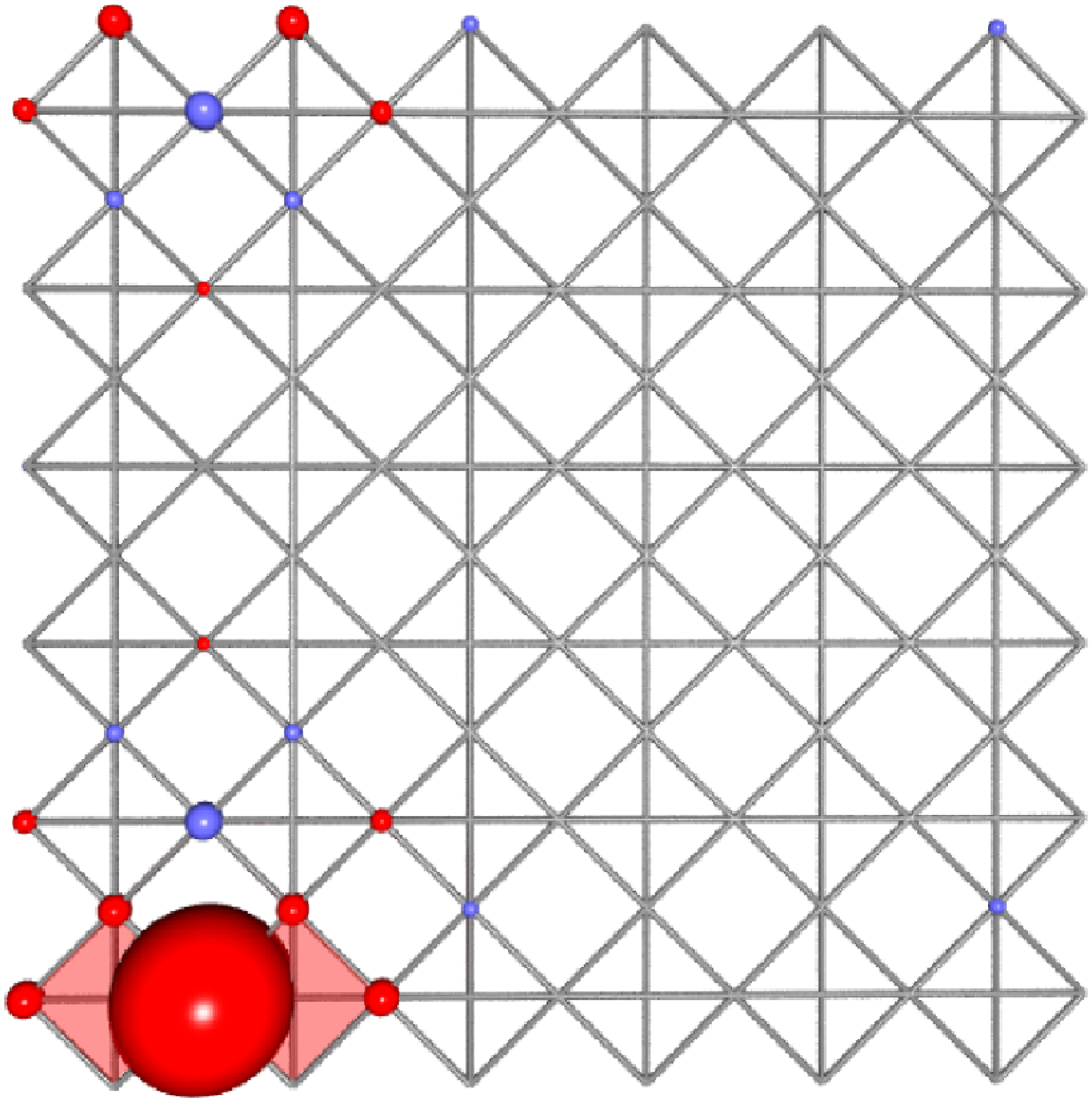}&
(e)\includegraphics[width=27mm]{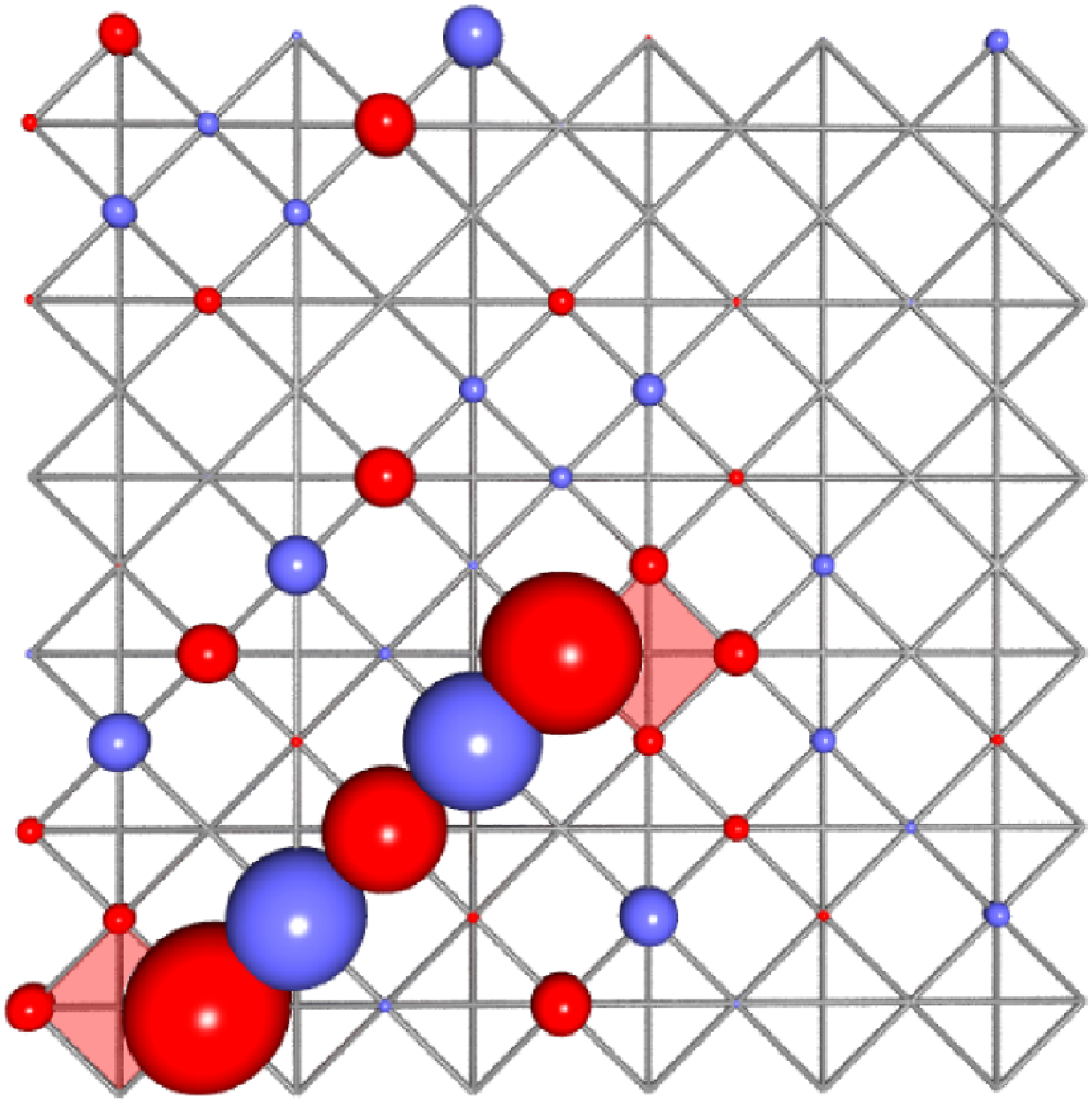}&
(f)\includegraphics[width=27mm]{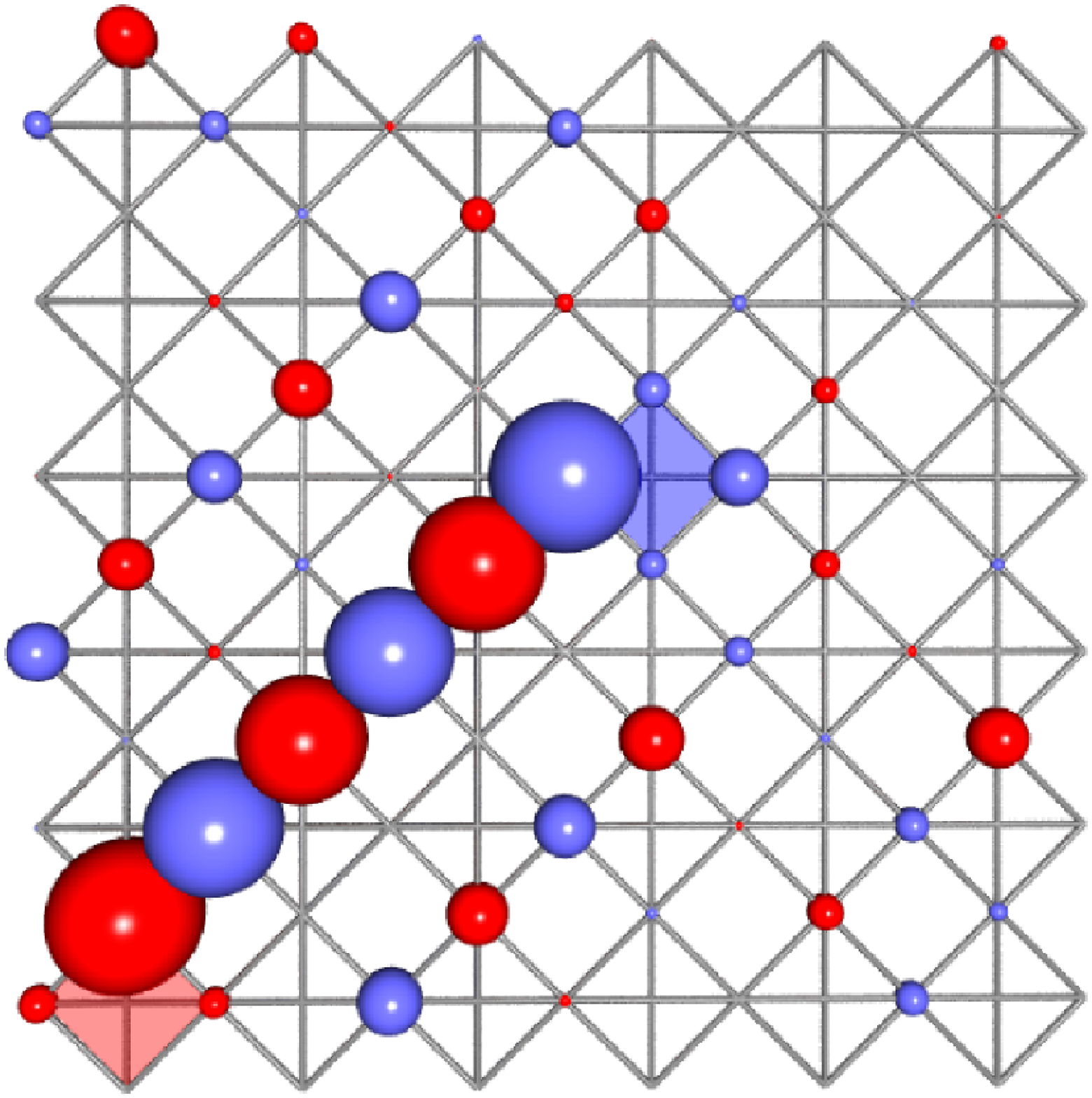}
\end{tabular}
\end{center}
\caption{(a)--(c) Local loss of kinetic energy due to the separation
of two (static) fractionally charged defects, i.e., particles or holes (fcp's
marked by light red squares or fch's marked by dark blue squares). The radii of
the circles are proportional to the local energy loss. (d)--(f)
Red (blue) circles show an increase (decrease) of the local density
(vacuum polarization due to the two fcp's or the fcp-fch pair).}
\label{cap:C4_string}
\label{cap:C4_string_fluct}
\end{figure}

We are now in the position to discuss the problem of confinement of fractional
charges. It is obvious that for $\mu \ge g$, i.e., when the ground state energy
is $E = 0$ it does not matter how far two charges e/2 of an added particle are
separated from each other since the energy is always 4V. However that is
different for $\mu = 0$, the case of physical interest. The energy change can
be decomposed into local contributions $\epsilon_i$ which include all hexagon
hopping processes involving site $i$. As seen in FIG. \ref{cap:C4_string}(a--c)
$\epsilon_i$ increases along the string connection between the two fractional
charges (compare with FIG. \ref{cap:C2_added}). This is plausible since ring
hopping requires alternating empty and occupied sites. In the vicinity of a
string of occupied sites these processes are reduced and $\epsilon_i$
increases. The same holds true when a pair $e/2$, $-e/2$ is
separated. Therefore in both cases the fractional charges are confined by a
constant confining force, similarly as quarks
[\onlinecite{gross73,politzer73}]. The energy increase is linear with charge
separation. From the numerical results we can deduce an energy increase of
$\Delta E \approx 0.2 |g|r$ where $r$ is the distance of the fractional charges
in units of the site distance $a$. The energy increase corresponds to a string
tension $T = 0.2 |g|$. It should be noticed that the restoring force is
weak. Assume that $V = 10t$ in which case $g = 0.12$ and $\Delta E = 0.024
ta$. The two fractionally charged particles $e/2$ will form a bound
state with a radius of order 50 - 100 lattice distances. Thus as soon as there
is a small doping concentration the average distance between the fractional
charges will be smaller than the diameter of the pairs and a plasma will
form. Note that when $r$ exceeds a critical value $r_c$ so that $\Delta E
= 0.2 g \cdot r_c$ then it is favourable to create fractionally charged
particle-hole pairs $e/2$, $-e/2$ in order to reduce a further
energy increase with $r$. Again, this resembles the generations of
quark-antiquark pairs in form of pions, when a quark is separated too
far from the other ones. 

One would like to know how fractionalization of charge could show up in
experiments. One quantity which depends strongly on it is the spectral function
$A({\bf k}, \omega) = A^+({\bf k}, \omega) + A^-({\bf k}, \omega)$ which is
defined through 

\begin{eqnarray}
A^{+}(\mathbf{k},\omega) & = &
\lim_{\eta\rightarrow0^{+}} - \frac{1}{\pi} \mbox{Im} \left\langle \psi_{0}^{N}
\left| c_{\mathbf{k}}\ \frac{1}{\omega + i\eta + E_{0}-H}\
c_{\mathbf{k}}^{\dag} \right|\psi_{0}^{N} \right\rangle \nonumber \\[2ex] 
A^{-}(\mathbf{k},\omega) & = & \lim_{\eta\rightarrow0^{+}} - \frac{1}{\pi}
\mbox{Im} \left\langle \psi_{0}^{N} \left|\  c_{\mathbf{k}}^{\dag}\
\frac{1}{\omega + i\eta - E_{0}+H}\  c_{\mathbf{k}} \right| \psi_{0}^{N}
\right\rangle~~~. 
\label{eq:15}
\end{eqnarray}

\noindent Here $| \psi_0^N \rangle$ is the ground state with energy $E_0$ of
the half-filled lattice. We have calculated $A({\bf k}, \omega)$ by using the
effective $H_{t-g}$ Hamiltonian for a finite cluster and different values of
$g$ (see FIG. \ref{cap:Spectral-density_gg}). For small values of $g$ the
diameter of the bound pair ($e/2$, $e/2$) formed by an added
particle exceeds the cluster size. Therefore the fractional charges appear to
be deconfined and as seen in FIG. \ref{cap:Spectral-density_gg}(a) there is no
quasiparticle peak in $A({\bf k}, \omega)$. This is different when $g$ is
sufficiently large, e.g., for $g = 1$. Here the bound pair remains within the
finite cluster and therefore a quasiparticle peak does appear (see
FIG. \ref{cap:Spectral-density_gg}(b). The internal degrees of freedom of the
bound ($e/2$, $e/2$) pair yield additional structure in $A({\bf
  k}, \omega)$. Note that in FIG. \ref{cap:Spectral-density_gg}(a) as well as
\ref{cap:Spectral-density_gg}(b) the bandwidth of $A({\bf k}, \omega)$ is nearly
twice as large than as without charge fractionalization in which case it would
by 8t. 

Hand in hand with the separation of two fractional charges goes a polarization
of the vacuum between the two objects. The modifications of the vacuum in the
vicinity of the string connecting the two fractional charges is shown in
FIG. \ref{cap:C4_string}(d--f). We may consider the vacuum polarization as
origin of the confining force.

The above considerations apply to the checkerboard lattice. As far as the
pyrochlore lattice is concerned we do not yet know whether fractionally charged
particles are confined or deconfined. There are numerical indications that the
charges are deconfined here due to the smaller effect of a string of occupied
sites on ring hopping. But only detailed numerical work can provide a solid
answer [\onlinecite{sikora08}].


\section{Strings and their time evolution}

\label{Sect:StringsTimeEv}

\begin{figure}[thb]
\includegraphics[height=57mm]{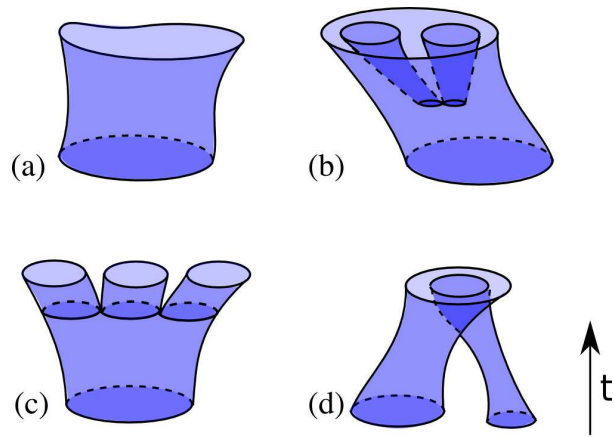}
\caption{Continuous representation of loop dynamics due to H$_{\rm eff}$. 
(a) Time evolution of loops due to B processes which conserve the topology.
(b) - (d) The same for A processes where H$_{\rm eff}$ induces three kinds of
  topological changes [\onlinecite{pollmann07d}].
\label{cap:Fig3}}
\end{figure}

As it was shown above, when spinless fermions occupy a checkerboard lattice at
half filling the strong correlations result in a complete loop covering of the
plane. The time evolution of these loops results in world sheets which 
replace world lines of single particle propagation. Therefore we are dealing
here with a simple form of a string theory, a very active and challenging topic
of present days field theory [\onlinecite{pollmann07b,ZwiebachBook}]. It is
interesting to consider the loop dynamics due to $H_{\rm eff}$ in a continuum
limit. This is shown in FIG. \ref{cap:Fig3}. There are processes denoted by B
in Section \ref{Sect:DynProc} which change only the shapes of the loops continuously. They
are visualized in FIG. \ref{cap:Fig3}(a). In distinction A processes in $H_{\rm eff}$ change the topology of a given loop covering. They were shown in
FIG. \ref{cap:C4_loops}(a) and give raise to a time evolution shown in
FIG. \ref{cap:Fig3}(b-d). 

\begin{figure}[thb]
\begin{center}
\begin{tabular}{cc}
(a)~\includegraphics[height=3cm]{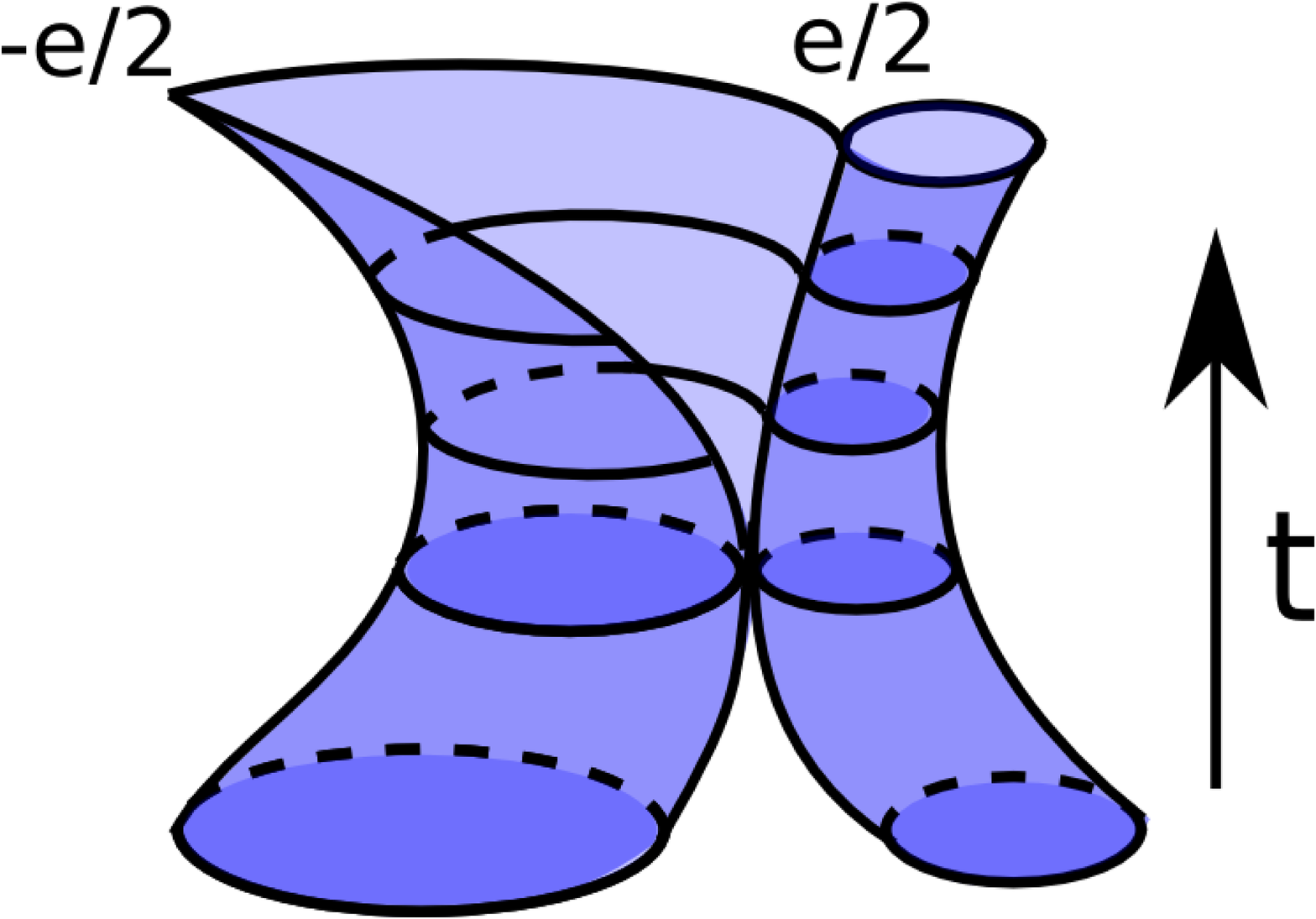}&
(b)~\includegraphics[height=3cm]{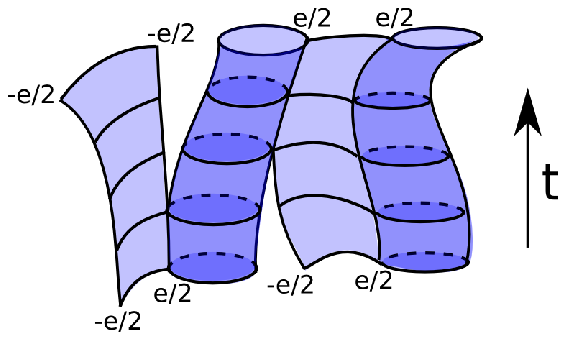}
\end{tabular}
\end{center}
\caption{When an energy $\Delta E > V$ is added to the system a loop is
    broken and an open string is generated. Thereby one end of the string is
    touching a closed loop.
(a) Time evolution in a continuous representation.
(b) Time evolution of two $\left( e/2, -e/2 \right)$ pairs and
    formation of $e$ and $-e$ particles.
\label{cap:Fig4}}
\end{figure}

When an energy $\Delta E > V$ is added to the ground state a loop can break up
and a pair of charges $e/2$, $-e/2$ is generated. 
This process is
shown in FIG. \ref{cap:Fig4}(a). When two or more pairs are generated new
combinations $e/2$, $e/2$ and $-e/2$, $-e/2$ may
form. They correspond to the creation of new particles out of the vacuum
(ground state). Note that in order that, e.g., the two fractional charges
$e/2$, $e/2$ combine to a particle with charge $e$ they must be
situated on different sublattices. This is schematically shown in FIG.
\ref{cap:Fig4}(b). Fractionally charged particles are always situated at the
end of open strings. In case of $e/2$ (but not $-e/2$) the open
end is always touching a closed loop. Similar considerations hold for
pyrochlore lattices at half fillings.


\section{Derivation of a gauge theory}

\label{Sect:DerGauTheo}

The problem of confinement vs. deconfinement of fractional charges on a
checkerboard lattice was described in Sect. \ref{Sect:ConfCeconf}. It can also
be discussed in terms of a field theory, more specifically a gauge
theory. Since the effective Hamiltonian conserves the number of particles on a
crisscrossed square the theory is invariant with respect to gauge changes on
those squares [\onlinecite{pollmann07b}]. This enables us to rewrite the
Hamiltonian in form of a U(1) lattice gauge theory and to compare it with that
of other models. Similar theories were derived before for the quantum-dimer
model [\onlinecite{Fradkin91,fradkin90}], for a three dimensional spin system
[\onlinecite{Hermele04,bergman06}] and for the ice model
[\onlinecite{castroneta04}]. The derivation follows the one given by Polyakov
[\onlinecite{Polyakovbook,Polyakov77}] for compact quantum electrodynamics. By
connecting the centers of crisscrossed squares we obtain a square lattice with
the particles sitting on links instead of sites. Define for each link ${\bf x},
{\bf x} + {\bf \hat e}_j$ between neighboring lattice sites ${\bf x}$ and ${\bf
  x} + {\bf \hat e}_j$ (j = 1,2) a variable ${\hat n}_j ({\bf x})$ with integer
eigenvalues. In order to express the effective Hamiltonian in terms of ${\hat
  n}_j ({\bf x})$ we first introduce its canonical conjugate, i.e., the phase
$\hat\phi_j ({\bf x}) \in [0, \pi]$. We note that exp. $[\pm i
  \tilde\phi_j ({\bf x})]$ acts like a ladder operator. In rewriting ring
hopping in terms of ${\hat n}_j ({\bf x})$ and $\hat\phi_j ({\bf x})$ we must
ensure that a link is either occupied by one particle or unoccupied. That leads
to the following form of the effective Hamiltonian (\ref{eq:C4_H_eff})

\begin{eqnarray}
H_{\rm eff} = \lim_{U \rightarrow \infty} U \sum_{{\bf x}_j} \left( \hat n_j
\left( {\bf x} \right) - \frac{1}{4} \right) + 2g \sum_{{\bf x}_j} \cos \left[
  \sum_{\left\{ \smallrech \smallrecv \right\}} \pm \hat\phi 	\right]  
\label{eq:16}
\end{eqnarray}

\noindent where the first term limits the eigenvalues of ${\hat n}_j ({\bf x})$
to 0 and 1 while the second term describes ring hopping. Note that here hexagon
ring hopping goes over into hopping around two neighboring squares (double
plaquettes) and that the signs of the phases alternate around the polygons. We
have assumed that by means of (\ref{eq:10}) the sign change in
(\ref{eq:C4_H_eff}) has been removed.  

Next we introduce staggered gauge and electric fields on the biparticle lattice  
\begin{eqnarray}
\hat A_j \left( {\bf x} \right) & = & \left( -1 \right)^{x_1 + x_2} \hat \phi_j
\left( {\bf x} \right) \nonumber \\
\hat E_j \left( {\bf x} \right) & = & \left( -1 \right)^{x_1 + x_2} \left( \hat
n_j \left( {\bf x} \right) - \frac{1}{2} \right)~~~.
\label{eq17}
\end{eqnarray}

\noindent The theory must incorporate the constraint that each lattice site is
touched by exactly two occupied links. This follows from the rule that each
tetrahedron contains two particles. Here this constraint reads

\begin{equation}
\left( \Delta_j \hat E_j \left( {\bf x} \right) - \rho \left( {\bf x} \right)
\right) \mid {\rm Phys} \rangle = 0
\label{eq18}
\end{equation}

\noindent with the lattice divergence defined by

\begin{equation}
\Delta_j \hat E_j \left( {\bf x} \right) = \hat E_1 \left( {\bf x} \right) -
\hat E_1 \left( {\bf x}-{\bf e}_1 \right) + E_2 \left( {\bf x} \right) - \hat
E_2 \left( {\bf x}-{\bf e}_2 \right)~~~.
\label{eq19}
\end{equation}

\noindent It is noticed that the constraint has here the form of Gauss' law. In
terms of $\hat E_j ({\bf x})$ and $\hat A_j ({\bf x})$ the effective
Hamiltonian (\ref{eq:16}) becomes 

\begin{eqnarray}
H_{\rm eff} & = & \lim_{U \rightarrow \infty} U \sum_{{\bf x}j} \left( \hat
E_j^2 \left( {\bf x} \right) - \frac{1}{4} \right) - 2g \sum_{{\bf x}j} \cos
\left( \sum_{\Box} \hat A_\ell \left( {\bf x} \right) + \sum_{\Box} \hat A_\ell
\left( {\bf x} - {\bf \hat e}_j \right) \right)~~~.   
\label{eq20}
\end{eqnarray}

\noindent The oriented sum of the vector potential around one plaquette is
\begin{equation}
\sum\limits_{\Box} \hat A_\ell ({\bf x}) = \hat A_1 ({\bf x}) - \hat A_1 ({\bf
  x} + {\bf e}_2) - \hat A_2 ({\bf x}) + \hat A_2 ({\bf x} + {\bf e}_1).
 \end{equation}
The Hamiltonian resembles the one of compact quantum electrodynamics in 2+1
dimensions [\onlinecite{Polyakovbook,Polyakov77}] but is not identical with
it. It is interesting that in the latter case two charges are confined and that
the energy grows linearly with there distance as it is the case in our
model. However, despite of the above mentioned resemblance the physical origin
of confinement is different in the two cases. As it turns out the U(1) gauge
theory formulation of our model system is gratifying but does not bring much
additional insight.


\section{Ferromagnetism generated by kinetic processes}

\label{Sect:Ferromagn}

The most common origin of ferromagnetism is spin exchange between
electrons. The latter may belong either to different atomic sites or to the
same site where intra-atomic exchange is the origin of Hund's rules. Pauli's
principle forbids electrons with parallel spins to come too close to each other
and reduces this way the mutual Coulomb repulsion. This should be compared with
superexchange, the standard mechanism for antiferromagnetism. Here it is the
kinetic energy which is optimized in the magnetic phase. Hence it is in general
a competition between an optimization of the Coulomb repulsion and of the
kinetic energy which favours magnetic order. In passing we mention other
sources of ferromagnetism such as RKKY interactions in metals or double
exchange, to name of few.

\begin{figure}[thb]
\begin{centering}
\begin{tabular}{ccc}
\tabularnewline
(a)\includegraphics[width=28mm]{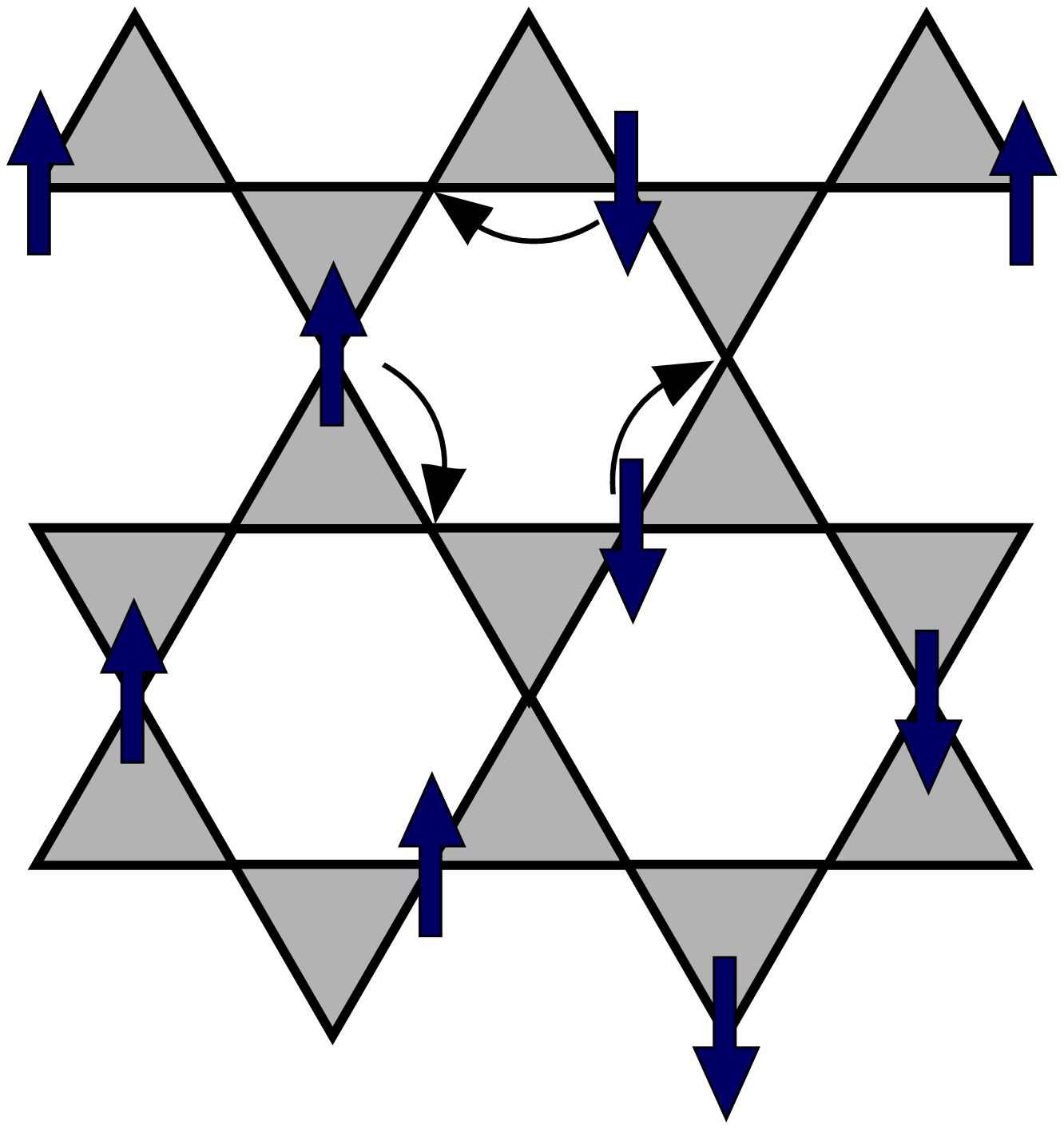}
(b)\includegraphics[width=28mm]{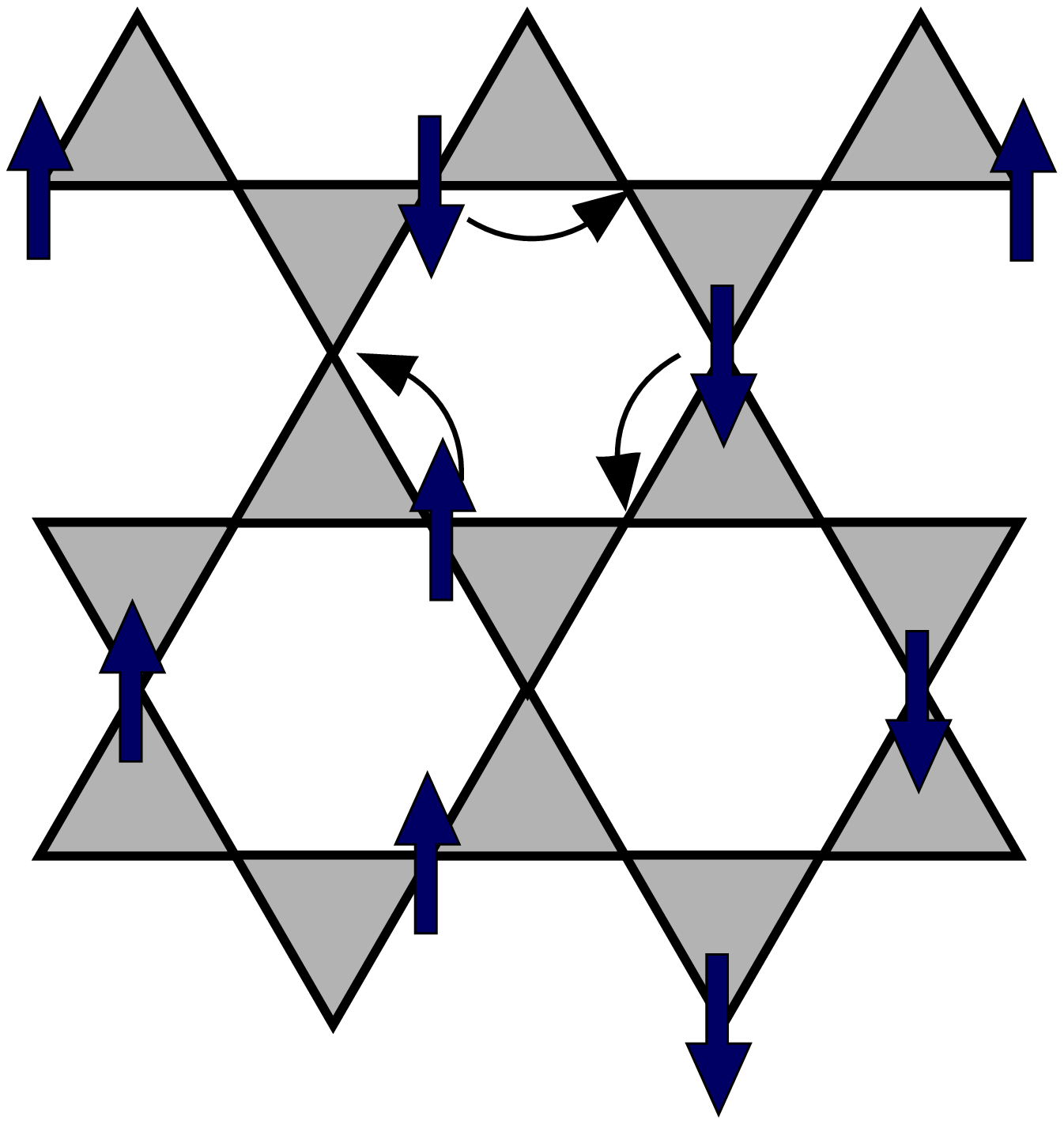}
\tabularnewline
(c)\includegraphics[width=28mm]{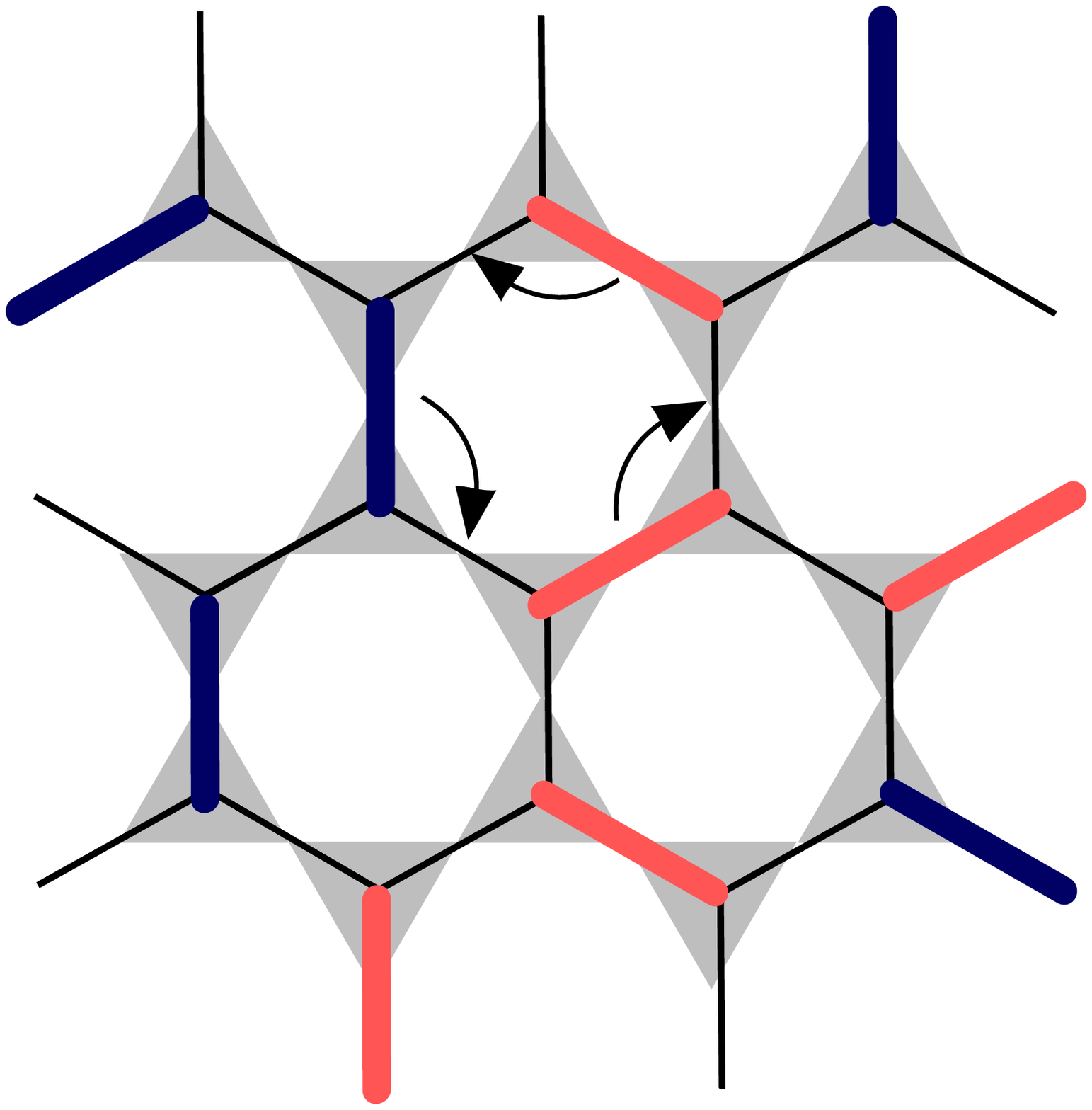}
(d)\includegraphics[width=28mm]{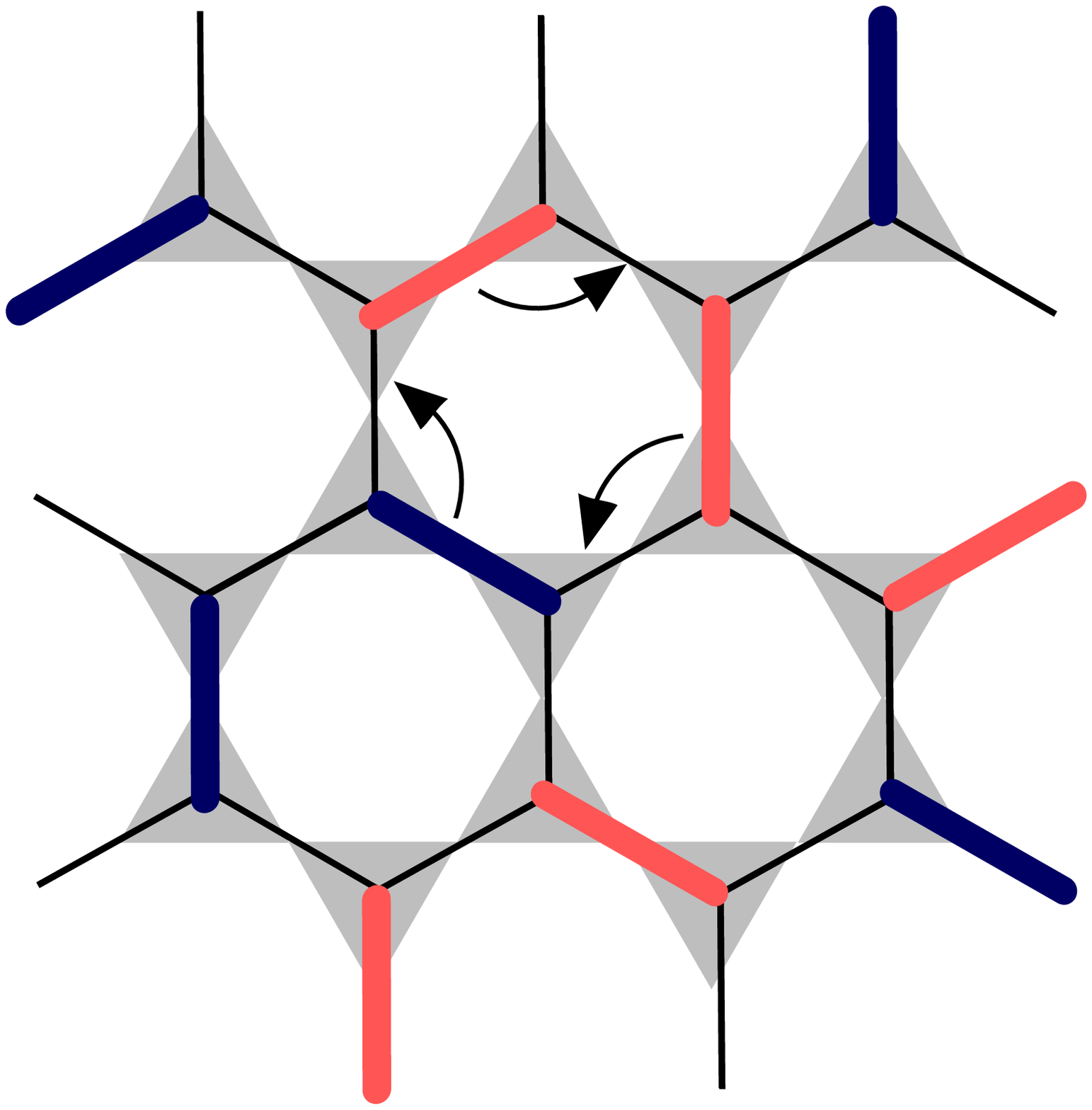}
\tabularnewline
\end{tabular}\par
\end{centering}
\caption{Panels (a) and (b) show two different allowed configurations on a
  kagome lattice which
fulfill the constraint of zero or one electron per site and one electron
of arbitrary spin per triangle. The arrows indicate possible ring-hopping
processes. All allowed configurations can be represented by colored
dimer coverings on a honeycomb lattice shown in (c) and (d) which is obtained
by connecting the centers of triangles. A blue dimer correspond to a spin up
and a red dimer to a spin down particle. 
\label{dimerpanel}}
\end{figure}

It might appear surprising that ferromagnetism may also be caused by purely
kinetic processes. We know of two specific examples where this is the case. One
is the ferromagnetic ground state discovered by Nagaoka
[\onlinecite{nagaoka66}]. It is due to a single hole moving in an otherwise
half-filled Hubbard system in the limit of infinite on-site repulsion $U$. The
proof is based in a theorem due to Perron and Frobenius. That
theorem states (see, e.g., Ref. [\onlinecite{saldanha95}])~ the largest
eigenvalue of a symmetric $n \cdot n$ matrix with only positive matrix elements
is positive and non-degenerate, while the corresponding eigenvector is
nodeless, i.e., we may choose it to have only positive components. The theorem
applies only to systems with a finite-dimensional Hilbert space. The same theorem is the basis of three-particle
ring exchange in $^3$He. In both cases ferromagnetism results from the 
motion of fermions because the ground state wavefunction is the
smoothest in this case (it is nodeless) and has the lowest kinetic energy. Here
we want to point out another source of ferromagnetism based on purely kinetic
effects. We demonstrate it by considering a partially filled kagome lattice
with electrons described by an extended Hubbard Hamiltonian (\ref{eq:4}). We
focus on the case of 1/6 filling implying one electron per triangle. In the
limit of $U \rightarrow \infty$ double occupancies of sites are excluded and
strong correlations, i.e., $|t| \ll V$ are assumed. When t = 0 the ground state
is macroscopically degenerate since all configurations with one electron of
arbitrary spin orientation per triangle are ground states. For examples see
FIG. \ref{dimerpanel}(a,b). By connecting the centers of the triangles of the
kagome lattice we obtain as the medial lattice a honeycomb lattice. Here
electrons sit on links instead of lattice sites. The different ground-state
configurations correspond here to two-colored (spin) dimer configurations (see
FIG. \ref{dimerpanel}(c,d). They are orthogonal because possible wavefunction
overlaps are neglected.

\begin{figure}[thb]
\begin{center}
\begin{tabular}{cc}
  \includegraphics[height=6cm]{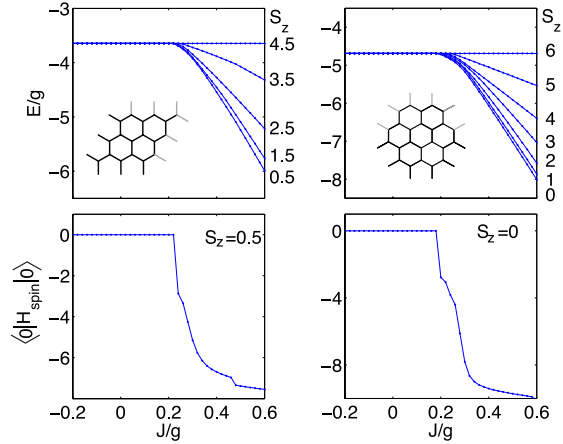}
\end{tabular}
\end{center}
\caption{Exact diagonalization of the two-color dimer model on 24-site
  honeycomb cluster. The upper panels show the ground-state energies of
  different $S_z$ sectors as a function of next-nearest neighbor coupling
  $J/g$. The lower ones show the expectation values of the spin part of the
  Hamiltonian. The ground state is denoted by $|0\rangle$
  [\onlinecite{pollmann07e}]. 
\label{cap:Both}}
\end{figure}

When we allow for $t \neq 0$ the ground-state degeneracy is lifted. To lowest
order of degenerate perturbations theory the effective Hamiltonian acting on
the subspace of configurations with precisely one dimer touching each site is
written as 

\begin{equation}
H_{\text{hex}}=-g\sum_{\left\{ \smallhexh\right\}
\left\{\blacktriangle\blacksquare\bullet\right\}}
\Big(\big|\hexaa\big\rangle\big\langle\hexb\big|
+\big|\hexab\big\rangle\big\langle\hexb\big|+\text{H.c.}\Big)
\label{eq:Hhex}
\end{equation}

\noindent where $g = 6t^3/V^2$. The sum is overall hexagon and spin
combinations symbolized by the three marks. Note that $H_{\rm hex}$ causes no
fermionic sign problem. If we would ignore the spin the model would go over
into the quantum dimer model (QDM) studied in
Ref. [\onlinecite{moessner01b}]. The ground state of the QDM is known to be
three-fold degenerate in the thermodynamic limit. It corresponds to the
valence-bond solid (VBS) plaquette phase with broken translational
symmetry. Here we are interested in spin correlations and notice that $H_{\rm
  hex}$ conserves both the total spin $S$ as well as its component
$S_z$. Without proof which is found in Ref. [\onlinecite{pollmann07e}] we state
that the Perron-Frobenius Theorem is applicable to the present system. 

Returning to $H_{\rm hex}$ we remark that by a simple gauge transformation the
sign of the plaquette flip in Eq. (\ref{eq:Hhex}) can be always chosen
negative, irrespective of the sign of $t$ [\onlinecite{pollmann07e}]. We
therefore choose all off-diagonal matrix elements of $H_{\rm hex}$ to be
non-positive. But before applying the Perron-Frobenius Theorem we have to
discuss the problem of ergodicity. The Hilbert space under consideration is
broken into different sectors corresponding to different $S_z$ which are not
connected by  $H_{\rm hex}$. Therefore each sector must be considered
separately. For the $S_{\rm tot}^z = S_{\rm max} = N_e/2$ sector, where $N_e$
is the number of electrons the ground state is unique and fully spin
polarized. The situation is more complex for the other sectors with $S_{\rm
  tot}^z \neq S_{\rm max}$. Although it can be shown that a fully spin
polarized state is also a ground state in those sectors, it has not been
possible to prove convincingly ergodicity, i.e., that the fully spin-polarized
ground state  is not only one among others. However, the
numerical studies on clusters strongly suggest that the ground state is indeed
unique. 

In order to study the robustness of the kinetic ferromagnetism, we introduce an
additional next-nearest neighbor antiferromagnetic interaction so that the
total effective Hamiltonian becomes $H_{\rm eff} = H_{\rm hex} + H_{\rm spin}$
with 

\begin{equation}
H_{\rm spin} = J \sum_{\ll ij \gg} \left( {\bf S}_i {\bf S}_j - \frac{1}{4} n_i
n_j \right)~~~. 
\label{eq21}
\end{equation}

\noindent Numerical diagonalization of clusters up to 24 sites show that the
ground state remains fully spin polarized up to values of $J/g < (J/g)_c
\approx 0.2$. This is shown in FIG. \ref{cap:Both} which demonstrates a
considerable robustness of kinetic ferromagnetism.

A ferromagnetic ground state is also found for a filling factor of 1/3. The
origin of kinetic ferromagnetism in the present examples differs from the ring
exchange proposed by Thouless for $^3He$ [\onlinecite{Thouless65}] insofar as
there the particles remain at their original location when they cyclically
permute while here they actually move around. It differs also from the
so-called flat band ferromagnetism of Mielke which he predicted for a kagome
lattice with fillings between 5/6 and 11/12
[\onlinecite{mielke91,mielke92}]. In his case any value of $U > 0$ is
sufficient and $V = 0$ while in our case $U \rightarrow \infty$ and $V/t$ is
large. In Mielke's case a completely flat band in a single particle description
of electrons on a kagome lattice influences so strongly the Stoner criterion
for the occurrence of ferromagnetism that any value of $U >0$ leads to a
ferromagnetic instability.


%
\bibliography{ReferencesPT}

\end{document}